\documentclass[%
 reprint,
superscriptaddress,
 amsmath,amssymb,
 aps,
prb,
]{revtex4-2}

\usepackage{graphicx}
\usepackage{dcolumn}
\usepackage{bm}
\usepackage{url}
\usepackage{array}
\usepackage{longtable}

\usepackage{xcolor}
\usepackage{soul}
\usepackage{float}

\begin{document}

\preprint{}

\title{Effect of pressure on the superconducting properties of Au substituted PdTe$_2$\\ with the CdI$_2$-type structure}

\author{Ayako Ohmura}%
 \email[E-mail: ]{ohmura@phys.sc.niigata-u.ac.jp}%
\affiliation{Faculty of Science, Niigata University, 8050, Ikarashi 2-no-cho, Nishi-ku, Niigata, 950-2181, Japan}%
\affiliation{National Institute for Materials Science (NIMS), Sengen 1-2-1, Tsukuba, Ibaraki 305-0047, Japan}%
\author{Kazuki Ichikawa}%
\affiliation{Graduate School of Science and Technology, Niigata University, 8050, Ikarashi 2-no-cho, Nishi-ku, Niigata, 950-2181, Japan}%
\affiliation{National Institute for Materials Science (NIMS), Sengen 1-2-1, Tsukuba, Ibaraki 305-0047, Japan}%
\author{Kyohei Tanaka}%
\affiliation{Graduate School of Science and Technology, Niigata University, 8050, Ikarashi 2-no-cho, Nishi-ku, Niigata, 950-2181, Japan}%
\author{Takashi Naka}%
\affiliation{National Institute for Materials Science (NIMS), Sengen 1-2-1, Tsukuba, Ibaraki 305-0047, Japan}%
\affiliation{Faculty of Science, University of the Ryukyus, 1 Senbaru, Nishihara-cho, Nakagami-gun, Okinawa, 903-0213, Japan}%
\author{Motoharu Imai}%
\affiliation{National Institute for Materials Science (NIMS), Sengen 1-2-1, Tsukuba, Ibaraki 305-0047, Japan}%
\affiliation{Faculty of Science, Tokyo Metropolitan University, 1-1 Minami-Osawa, Hachioji, Tokyo 192-0397, Japan}%
\author{Fumihiro Ishikawa}%
\affiliation{Faculty of Science, Niigata University, 8050, Ikarashi 2-no-cho, Nishi-ku, Niigata, 950-2181, Japan}%
\author{Takayuki Nakane}%
\affiliation{National Institute for Materials Science (NIMS), Sengen 1-2-1, Tsukuba, Ibaraki 305-0047, Japan}%
\author{Anne de Visser}%
\affiliation{Van der Waals-Zeeman Institute, University of Amsterdam, Science Park 904, 1098 XH Amsterdam, The Netherlands}%

\date{\today}

\begin{abstract}
Transition metal ditellurides with the CdI$_2$-type structure are materials with intriguing superconducting and electronic properties as demonstrated by PdTe$_2$. Gold substituted PdTe$_2$, Au$_x$Pd$_{{\rm 1}-x}$Te$_2$, adopts the CdI$_2$-type structure for a Pd content larger than 43 at.\% at room temperature, and in this range enhanced superconductivity with a critical temperature ($T_{\rm c}$) above 4 K has been reported (Kudo \textit{et al}., PRB {\bf 93}, 140505, 2016). Here we present the effect of pressure on the structural and superconducting properties of Au$_x$Pd$_{{\rm 1}-x}$Te$_2$ for $x=0.15$, 0.25 and 0.35 with $T_{\rm c} =2.7$, 4.1, and 4.6 K at 1 atm, respectively. Synchrotron radiation x-ray diffraction shows that the CdI$_2$-type structure remains stable up to 8 GPa for all three compositions and that they have almost the same volume compressibility. Heat capacity measurements show that Au substituted PdTe$_2$ exhibits type-II superconductivity, that evolves from weak-coupling BCS for $x=0.15$ to intermediate coupling for $x=0.25$ and 0.35.  Electrical resistivity measurements up to a pressure of 2.5 GPa reveal that $T_{\rm c}(P)$ for $x=0.25$ and 0.35 has a shallow maximum with $T_{\rm c}^{\rm max}$ is 4.2 and 4.7 K at $P\sim 0.3$ and 0.7 GPa, respectively, whereas a monotonic decrease is observed for $x=0.15$. Using the McMillan equation, with the Debye temperature ${\it \Theta}_{\rm R}$ extracted from the resistance data under pressure as input, we find a smooth variation of $T_{\rm c}$ with the electron-phonon coupling strength $\lambda_{\rm ep}$ spanning the entire pressure and concentration range, from PdTe$_2$ to 35 at.\% Au substitution. We conclude superconductivity across the Au$_x$Pd$_{{\rm 1}-x}$Te$_2$ series under pressure is governed by a delicate balance between lattice stiffening and the strength of the electron phonon coupling. 


\end{abstract}


\maketitle


\section{\label{sec:level1}Introduction}   

Transition metal ditellurides with the CdI$_2$-type structure, $M$Te$_2$ ($M$ = Ni, Pd, Pt, Ir), attract much attention because of their exotic properties. They are classified as type-II Dirac semimetals, which have a tilted Dirac cone and a topologically non-trivial surface state \cite{Soluyanov2015, Huang2016, Yan2017, Bahramy2018, Xu2018, Ghosh2019}. In particular, PdTe$_2$ has considerable potential as a topological superconductor, due to the coexistence of superconductivity with a transition temperature $T_{\rm c}\sim 1.6$ K at ambient pressure \cite{Guggenheim1961} and the topological electronic structure attributed to the CdI$_2$-type crystal structure \cite{Bahramy2018, Fei2017, Noh2017, Clark2018, Das2018, Amit2018_2}.  

Much research has been carried out on the superconducting properties of PdTe$_2$ at ambient pressure \cite{Leng2017, Amit2018, Salis2018, Teknowijoyo2018, Kim2018, Le2019, Voerman2019, Sirohi2019, Leng2019, Salis2021, Erik2019, Anemone2021}. PdTe$_2$ is a rare example of a binary compound showing type-I superconductivity with $T_{\rm c}$ = 1.64 K and a critical field $\mu_0 H_{\rm c}$(0) = 13.6 mT \cite{Leng2017}. Bulk superconductivity has been revealed to be conventional with a full superconducting gap confirmed by heat capacity \cite{Amit2018, Salis2021} and penetration depth measurements \cite{Salis2018, Teknowijoyo2018}. Additionally, Leng {\it et al}. observed an unusual type of surface superconductivity in measurements of the ac-susceptibility, with a transition temperature $T_{\rm c}^{\rm S}\sim$ 1.3 K and critical field $\mu_0 H_{\rm c}^{\rm S}$(0) = 34.9 mT. Hence an unusual superconducting phase diagram was obtained and it has been suggested that superconductivity of the surface sheet has a topological nature \cite{Leng2017}. We remark that Schimmel \textit{et al.} recently reported evidence for surface superconductivity by scanning tunneling microscopy/spectroscopy in a related material: the topological Weyl semimetal t-PtBi$_2$ \cite{Schimmel2024}.

Research on the effect of doping and substitution on PdTe$_2$ is progressing, and the following two findings have been reported \cite{Ryu2015, Kudo2016, Hooda2018, Salis2022}. Firstly, the enhancement of superconductivity. In Cu$_x$PdTe$_2$ \cite{Ryu2015}, the transition temperature is raised to $T_{\rm c}\sim$ 2.6 K for $x=0.05$. This is attributed to the hybridization between the Te-$p$ and Cu-$d$ orbitals in the Van der Waals gap, which enhances the carrier density at the Fermi level, $E_f$, and consequently increases $T_{\rm c}$. A larger increase of $T_{\rm c}$ can be created in the Au$_x$Pd$_{{\rm 1}-x}$Te$_2$ series \cite{Kudo2016}. Kudo {\it et al}. substituted Pd on the Au site in AuTe$_2$, which has a monoclinic $C2/m$ structure, and found that the crystal structure changes to trigonal $P$\={3}$m$1 (CdI$_2$-type) at about 43~at.\%~Pd at room temperature. This is attributed to the breaking of the Te-Te dimers. Additionally, Kudo \textit{et al}. propose that the increase of the density of states (DOS) at $E_f$ induces an increase in $T_{\rm c}$. In the CdI$_2$-type structure superconductivity emerges, with a maximum $T_{\rm c}$ of 4.65 K near the crystallographic phase boundary, which is located at 55~at.\%~Pd at liquid helium temperatures. The second important finding is the evolution from type-I to type-II superconductivity as confirmed by Cu-doping \cite{Ryu2015} and Pt-substitution \cite{Salis2022}. In particular, for the latter material small deviations from stoichiometry, {\it e.g.} Pd$_{0.97}$Pt$_{<0.004}$Te$_{2.03}$, result in type-II superconductivity. With regard to this point, Salis {\it et al}. \cite{Salis2022} discussed that PdTe$_2$ essentially has its superconducting instability close to the type-I/II border and that small amounts of disorder are sufficient for a transition to a superconductor of the second kind.

 Compared with chemical pressure achieved by doping or substitution, mechanical pressure, which continuously tunes the physical properties, has also been effective in the research on superconducting transition metal ditellurides \cite{Soulard2005, Xiao2017, Leng2020, Yang2021, Furue2021, MarcPhD2023, Kitagawa2013, Paris2016, Qi2020, Otsuki2021}. In our own work on PdTe$_2$ we showed that $T_{\rm c}(P)$ has a weak maximum near 1 GPa. This is primarily attributed to the pressure variation of the lattice stiffening rather than that of the carrier density \cite{Furue2021}. Moreover, the unusual $H$--$T$ phase diagram with bulk and surface superconductivity is robust under pressure \cite{Leng2020}. 
 
 
 Despite the broad interest in this group of materials, and transition metal dichalcogenides in general \cite{Hooda2021}, systematic experimental high pressure studies are scarce. In this respect the Au$_x$Pd$_{{\rm 1}-x}$Te$_2$ series offer an excellent opportunity. The trigonal CdI$_2$-type structure is stable in a wide composition range and $T_{\rm c}$ varies from 1.6 K for PdTe$_2$ to 4.6 K for 35~at.\%~Au substitution. Au substitution results among other things in carrier doping and affects the intralayer spacing. In our study we combine the effect of pressure on the structural properties, investigated by synchrotron x-ray diffraction, with pressure tuning of the superconducting state for $x=0.15$, 0.25 and 0.35. Superconductivity under pressure is probed by electrical resistance  measurements, and by magnetization and specific heat measurements at ambient pressure. Using the McMillan equation, with the Debye temperature ${\it \Theta}_{\rm R}$ extracted from the resistance under pressure data as input, we find a smooth variation of $T_{\rm c}$ with the electron-phonon coupling strength $\lambda_{\rm ep}$ over the entire Au concentration ($x \leq 0.35$) and pressure range. 
 

\section{Experimental and calculation details}

Polycrystalline Au$_x$Pd$_{{\rm 1}-x}$Te$_2$ samples with compositions $x=0.15$, 0.25 and 0.35 were prepared by solid-state synthesis as described in a previous study \cite{Kudo2016}. Stoichiometric mixtures of Au (4N), Pd (3N), and Te (5N) fine powders were ground and sealed in an evacuated quartz tube. The tube was heated at 500\,$^\circ$C for 24 h, and then its product was ground again and pelletized. The pellets were vacuum-sealed in a quartz tube and annealed twice at 400--600\,$^\circ$C for 24 h. The crystal structure and superconductivity of the final products at ambient pressure were identified by x-ray diffraction (XRD), and DC magnetization (DCM) and heat capacity (HC) measurements, respectively. For the DCM and HC measurements, a magnetic properties measurement system (MPMS, Quantum Design) and a physical properties measurement system (PPMS, Quantum Design) were used, respectively. The actual compositions were determined by Electron Probe Micro Analysis (EPMA) and are Au$_{0.13}$Pd$_{0.88}$Te$_{1.99}$ for $x=0.15$, Au$_{0.29}$Pd$_{0.74}$Te$_{1.97}$ for $x=0.25$ and Au$_{0.38}$Pd$_{0.66}$Te$_{1.96}$ for $x=0.35$. Details are presented in Sect. 1 of the Supplementary Material (SM). Throughout this paper we will refer to the different compositions by their nominal value. For $x$\,=\,0.35 we confirmed by Energy Dispersive x-ray Spectroscopy (EDX) that the final product is phase pure and homogeneous. The carrier density for this composition has been determined by a Hall effect measurement and amounts to $n=1.7\times 10^{23}$ cm$^{-3}$ (see Sect. 2 in the SM).

A diamond anvil cell (DAC) was used for synchrotron radiation XRD at room temperature under high pressure. The experimental conditions and preparation of the DAC and the powdered samples in this study are the same as reported for PdTe$_2$ \cite{Furue2021}. The XRD study under hydrostatic conditions was performed using a 4:1 mixture of methanol and ethanol as the pressure-transmitting medium. Synchrotron radiation XRD studies were carried out at the beamline AR-NE1A of the Photon Factory in the High-Energy Accelerator Research Organization (KEK) in Tsukuba, Japan. An incident beam was tuned to an energy of $\sim 29.7$ keV ($\lambda \sim 0.417$ \AA) with a cross-section of 75$\times$75 $\mu$m$^2$. All diffraction patterns were measured using two-dimensional detectors. The experimental pressures were determined using the ruby fluorescence method. The Rietveld analysis performed to optimize the structural parameters was made with the crystallographic program JANA2020 \cite{JANA2006}. Typical results of the analysis are shown in Fig.~\ref{fig:1} for $x=0.35$ at 0.25 GPa.

Electrical resistivity measurements under high pressure, at low temperature and in magnetic fields were performed with a hybrid CuBe/NiCrAl piston-cylinder cell in the PPMS. For each run, two experimental samples were sealed in a Teflon capsule together with Daphne oil 7373 as the pressure-transmitting medium. The pressure value generated in the capsule against the load on the high-pressure cell was calibrated by the superconducting transition temperature of lead as a reference standard \cite{Boughton1970}, which was measured separately by the ac-susceptibility method in advance; the generated pressure has an accuracy of $\pm 3\%$. The samples were compressed at room temperature and subsequently cooled to about 2 K. The electrical resistivity was measured using the AC resistance mode of the PPMS with for each sample a different frequency (18.3 and 21.6 Hz) and an excitation current of 1 mA. The temperature variation of the electrical resistivity was measured up to a pressure of $\sim$2.5 GPa in zero and applied magnetic ﬁelds.


DOS calculations were performed using the projector augmented wave (PAW) method, as implemented in the Vienna Ab initio Simulation Package (VASP) code within the generalized gradient approximation (GGA) of Perdew, Burke, and Ernzerhof (PBE) \cite{Kresse1996, Kresse1999}. The energy cutoff of 400 eV and the $\Gamma$-centered 9$\times$9$\times$10 $k$-meshes were used. The electronic convergence criterion was set to 10$^{-6}$ eV. The VASPKIT code was used to create the input files \cite{Wang2021}. The orbital-projected DOS (p-DOS) were calculated from the VASP output using the Local Orbital Basis Suite Towards Electronic Structure Reconstruction (LOBSTER) code \cite{Deringer2011, Maintz2016, Nelson2020}. The PAW functions were projected onto local orbitals using the basis-function set pbeVaspFit2015 with the autorotate option.

\section{Results}

In this section we concentrate on the results for the $x=0.35$ compound, which has the highest superconducting transition temperature. For the raw data and analysis of $x=0.15$ and 0.25 we refer to Sections 3-5 in the SM.

\subsection{Structural properties at ambient and high pressures}

\begin{figure}[t]
 \begin{center}
   \includegraphics[clip,width=8.5cm]{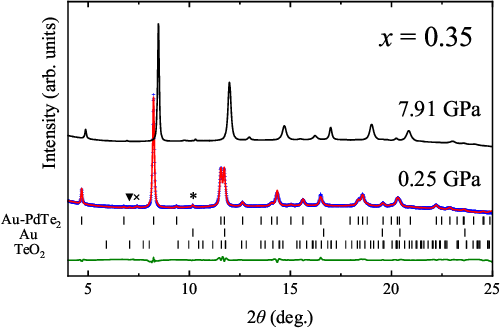}
  \caption{Diffraction patterns of Au$_{0.35}$Pd$_{0.65}$Te$_2$ measured at 0.25 and 7.91 GPa at room temperature, and the Rietveld refinement result of the lower pattern. Blue symbols (+), and red and green lines, represent the experimental diffraction intensity, the calculated intensity, and the residual error, respectively. The three sets of vertical bars below the diffraction patterns indicate the diffraction angles of the $P\overline{3}m$1, $Fm\overline{3}m$, and $P4_12_12$ structure for Au$_x$Pd$_{1-x}$Te$_2$, gold ($\ast$ in the pattern), and TeO$_2$ ($\blacktriangledown$), respectively. The cross symbol ($\times$) marks an unassigned peak. The goodness of fit for this pattern is 4.20.}
    \label{fig:1}
  \end{center}
\end{figure}

In the diffraction patterns of the Au substituted compounds most reflections can be indexed with the $P$\={3}$m$1 space group, but a few reflections from metallic Au and TeO$_2$ are also observed. Figure~\ref{fig:1} shows the x-ray diffraction patterns of $x=0.35$ measured at 0.25 and 7.91 GPa at room temperature, along with the Rietveld refinement results at 0.25 GPa. The diffraction pattern for $x=0.35$ was fitted using profiles of the three coexisting phases. However, an unassigned tiny peak, marked by a cross symbol ($\times$) cannot be accounted for. The Rietveld analysis without the unassigned peak results in volume fractions of $\sim97$\% for Au$_{0.35}$Pd$_{0.65}$Te$_2$, $\sim1$\% for gold, and $\sim2$\% for TeO$_2$, respectively. These impurity phases are not superconducting \cite{Hamilton2015, Gao2019} and their presence does not affect our analysis. Fig. S4 in the SM reveals that the diffraction peaks gradually shift to higher scattering angles with increasing pressure. For all compositions the diffraction profile remains the same and no new peaks appear up to $\sim$8 GPa. This leads us to conclude that the $P$\={3}$m$1 structure is stable up to the maximum pressure.

\begin{table}[b]
\caption{\label{tab:1} Lattice constants, $a_0$ (\AA) and $c_0$ (\AA), and unit cell volume, $V_0$ (\AA$^3$), at ambient pressure, and the bulk modulus, $B_0$ (GPa), and its pressure derivative, $B_0^{'}$, obtained by fitting the Birch-Murnaghan equation of state obtained for Au$_x$Pd$_{{\rm 1}-x}$Te$_2$ with $x=0.15$, 0.25 and 0.35. }
\begin{ruledtabular}
\begin{tabular}{cccccc}
\mbox{$x$}&\mbox{$a_0$}&\mbox{$c_0$}&\mbox{$V_0$}&\mbox{$B_0$}&\mbox{$B_0^{'}$}\\
\mbox{}&\mbox{(\AA)}&\mbox{(\AA)}&\mbox{(\AA$^3$)}&\mbox{(GPa)}&\mbox{}\\
\hline\\
0.15  & 4.0564(3) & 5.1418(8) & 73.263(8)  & 61(1)  & 7.5(4)    \\
0.25  &4.0680(5) & 5.1271(10) & 73.472(10)  & 63(2)  & 6.5(4)   \\
0.35 & 4.0871(4) & 5.1233(9)  & 74.100(9) & 58.6(9)  & 8.7(3)   \\
\end{tabular}
\end{ruledtabular}
\end{table}

\begin{figure}[htpb]
  \begin{center}
    \includegraphics[clip,width=7cm]{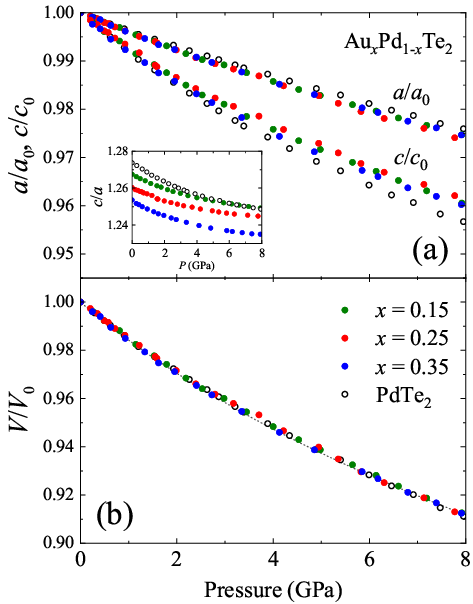}
    \caption{Pressure variation of the lattice constants and volume normalized to the values at ambient pressure for Au$_x$Pd$_{{\rm 1}-x}$Te$_2$: (a) $a/a_0$ and $c/c_0$ and (b) $V/V_0$. The inset of (a) shows the ratio $c/a$ of the lattice constants as a function of pressure. Closed green, red, and blue circles show values for $x=0.15$, 0.25 and 0.35, respectively. A broken line in (b) is the result of a fit to the BM-EOS for $x=0.35$ (see text). Open circles show those of PdTe$_2$, obtained from our previous work \cite{Furue2021}, for comparison.}
    \label{fig:2}
  \end{center}
\end{figure}

Table~\ref{tab:1} lists the lattice parameters at ambient pressure determined from the Rietveld analysis. The composition dependence of the $a_0$- and $c_0$-values has an inverse tendency, consistent with previous results \cite{Kudo2016}. An increase in the unit cell volume with $x$ is expected, since the atomic radius of Au is larger than that of Pd. The increase is mainly reflected in the elongation of the $a$-axis. In contrast, the shrinkage of the $c$-axis with increasing Au substitution is considered to be due to a special structural feature of AuTe$_2$, namely Te$_2$ dimers. The crystal structure of AuTe$_2$ includes a Te$_2$ dimer, and its average structure is of the monotonically distorted CdI$_2$-type with $C2/m$ symmetry \cite{Kudo2016, Kudo2013}. According to theoretical calculations by Streltsov \textit{et al.} \cite{Streltsov2018}, in AuTe$_2$ charge disproportionation occurs on the Au site. This induces the coexistence of AuTe$_4$ plaquettes (Au$^{3+}$) and AuTe$_2$ dumbbells (Au$^{1+}$), which in turn results in the dimerization of Te atoms. Therefore, as the amount of Au substitution in Au$_x$Pd$_{1-x}$Te$_2$ increases, a smaller Te-Te distance is preferred. This distance corresponds to the interlayer Te-Te spacing in the CdI$_2$-type structure. Since this bonding is oriented approximately along the $c$-axis it is consistent with the shrinkage of the $c$-axis.

Figure~\ref{fig:2} shows the pressure variation of the lattice constants and the unit cell volume normalized to their values at ambient pressure. The variations of $a/a_0$ and $c/c_0$ for $x=0.15$, 0.25 and 0.35 are almost identical. Compared to PdTe$_2$ \cite{Furue2021}, the $a$-axis of the doped compounds is more compressible. Inversely, the $c$-axis is less compressible. As regards the latter, the bond strength in the interlayer Te-Te spacing mentioned above probably makes it less compressible. In addition, PdTe$_2$ has a wider Van der Waals gap due to the larger $c/a$ ratio, which suggests that the compressibility along the $c$-axis is larger.

The pressure variation of the unit cell volume is consistent between the three compositions by canceling out the difference in $a/a_0$ and $c/c_0$. The compression curve can be fitted by the Birch-Murnaghan equation of state (BM-EOS) \cite{BMEOS1947}:
\begin{gather}
P=\frac{3}{2}B_0\left \{ \left( \frac {V_0}{V}\right)^{\frac{7}{3}}- \left (\frac {V_0}{V} \right)^{\frac{5}{3}} \right \} \cdot \nonumber\\ 
\left [ 1+\frac {3}{4}\left (B_0^{'}-4 \right ) \left \{  \left( \frac {V_0}{V} \right)^ {\frac{2}{3}}-1 \right \} \right], \nonumber
\end{gather}
where $V_0$ and $V$ are volumes at ambient and high pressures, $P$ is in units of GPa, $B_0$ is the bulk modulus, and $B^{'}_0$ is its pressure derivative. The fitted $B_0$ and $B^{'}_0$ parameters for each composition are listed in Table I. The $B_0$-value for $x=0.15$ and 0.25 is the same within the error bar, reflecting the compression curves. The bulk modulus of $x=0.35$ is smaller than for $x=0.15$ and 0.25, which can be attributed to the larger $B^{'}_0$, which depends on the slope of the initial compression curve.  


\subsection{Superconducting properties at ambient pressure}

\begin{figure}[htpb]
  \begin{center}
    \includegraphics[clip,width=6.5cm]{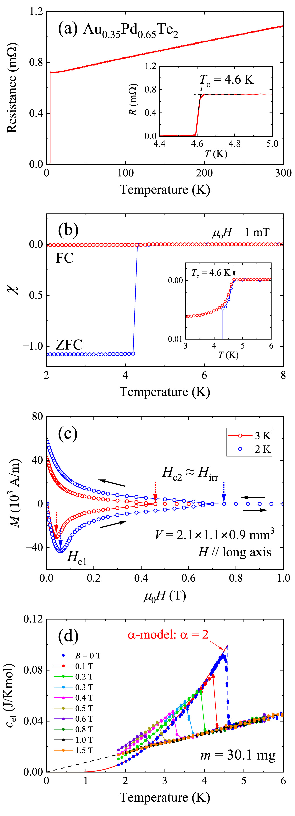}
    \caption{Ambient pressure experimental results of the superconducting transition in Au$_{0.35}$Pd$_{0.65}$Te$_2$: (a) electrical resistance as a function of temperature, (b) zero field-cooled (ZFC) and field-cooled (FC) dc susceptibility $\chi$ measured at a field of $\mu_0H=1$ mT, (c) DC magnetization per unit volume as a function of applied ﬁeld at temperatures of 2 and 3 K, and (d) temperature dependence of the electronic specific heat at magnetic fields from 0 to 1.5 T. The black dashed line shows the normal state $c_{\rm el}= \gamma T$. The red line shows $c_{\rm el}(T)$ of the superconducting state at zero field calculated with $\alpha=2$ \cite{Johnson2013} (see text). Insets in (a) and (b) show a magnified view around the onset temperature of the superconducting transition. In (c) the lower, $H_{\rm c1}$, and upper critical field, $H_{\rm c2}$, are indicated by vertical arrows. }
    \label{fig:3}
  \end{center}
\end{figure}

\begin{table*}[htpb]
\caption{\label{tab:2} Parameters derived from the heat capacity measurements at ambient pressure: the transition temperature $T_{\rm c}$ (K), the Sommerfeld coefficient $\gamma$ (mJ/mol\,K$^2$), the phononic coefficient $\beta$ (mJ/mol\,K$^4 $), the Debye temperature ${\it \Theta}_{\rm D}$ (K), the heat capacity jump at $T_{\rm c}$ $\Delta c/\gamma T_{\rm c}$, the $\alpha$-value, the upper critical field at 0 K $\mu_0 H_{\rm c2}$(0) (T), estimated from the WHH curves, and the coherence length $\xi$ (nm).}
\begin{ruledtabular}
\begin{tabular}{ccccccccc}
\mbox{$x$}&\mbox{$T_{\rm c}$}&\mbox{$\gamma$}&\mbox{$\beta$}&\mbox{${\it \Theta}_{\rm D}$}&\mbox{$\Delta c/\gamma T_{\rm c}$}&\mbox{$\alpha$}&\mbox{$\mu_0 H_{\rm c2}(0)$}&\mbox{$\xi$}\\
\mbox{}&\mbox{(K)}&\mbox{(mJ/mol\,K$^2$)}&\mbox{(mJ/mol\,K$^4$)}&\mbox{(K)}&\mbox{}&\mbox{}&\mbox{(T)}&\mbox{(nm)}\\
\hline\\
0.15  & 2.7 & 6.09  & 0.619 & 210 &1.43& 1.76&0.186&42.1 \\
0.25  &4.1 & 7.15  & 0.733 & 199 &1.71& 1.92&0.684&21.9  \\
0.35 & 4.6 & 7.65 & 0.890  & 187 &1.74& 1.94&1.02&18.0 \\
\end{tabular}
\end{ruledtabular}
\end{table*}

We performed resistivity, DC magnetization and heat capacity measurements to characterize the superconducting state of the Au substituted compounds at ambient pressure. In the resistivity measurement for $x=0.35$, shown in Fig.~\ref{fig:3}(a), metallic behavior is observed and a transition to a superconducting state occurs with an onset temperature of $T_{\rm c}=4.6$ K. Figure~\ref{fig:3}(b) shows the temperature variation of the DC susceptibility, $\chi$, in an applied magnetic field of $\mu_0H=1$ mT. The data were taken after cooling in zero field (ZFC) and field cooled (FC). The value of $\chi$ is corrected according to $\chi=\chi_{\rm m}/(1-N\chi_{\rm m})$, where $\chi_{\rm m}(=M/H)$ is the DC susceptibility obtained experimentally and $N$ is the demagnetization factor \cite{Chen2002}. A value of $N\sim0.19$ was estimated from the dimensions given in Fig.~\ref{fig:3}(c). While a significant diamagnetic shielding is observed in the ZFC data, its response in the FC data is tiny, indicating that magnetic flux is pinned without expulsion from the sample. The transition temperature $T_{\rm c}=4.6$ K can be estimated from the onset temperature in both ZFC and FC data and is consistent with that of the resistance data. 

The field variation of the magnetization in DC fields at temperatures of 2 and 3 K is shown in Fig.~\ref{fig:3}(c). Magnetic fields were applied along the long axis of the sample. The magnetization curves exhibit the features of a type-II superconductor with hysteresis. At $T=2$ K, the Meissner state is stable up to $\mu_0H=60$ mT, denoted as $H_{\rm c1}$ in Fig.~\ref{fig:3}(c), and subsequently the mixed state exists up to $\mu_0H=0.75$ T. We take $H_{\rm c2}\approx H_{\rm irr}$ as indicated by the arrow.

Figure~\ref{fig:3}(d) shows the temperature dependence of the electronic specific heat, $c_{\rm el}(T)$, measured at fields from $\mu_0H=0$ to 1.5 T in the temperature range of 1.8 to 6 K. To obtain the $c_{\rm el}(T)$-curve, we added a higher-order $\delta T^5$ term to the standard form $c=\gamma T + \beta T^3$, where $\gamma$ is the Sommerfeld coefficient and $\beta$ is the phononic coefficient; the inclusion of the $\delta T^5$ term improves the linear dependence of $c_{\rm el}(T)$ on temperature in the normal state. This resulted in the fitted values: 7.65 mJ/mol\,K$^2$ for $\gamma$, 0.890 mJ/mol\,K$^4$ for $\beta$, and 3.36$\times10^{-3}$ mJ/mol\,K$^6$ for $\delta$. The Debye temperature ${\it \Theta}_{\rm D}=187$ K is calculated from the relation ${\it \Theta}_{\rm D}=[(12\pi^4 SR)/(5\beta)]^{1/3}$, where $R$ is the molar gas constant and $S$ is the number of atoms per formula unit. The field variation of the $c_{\rm el}(T)$-curves shows the depression of superconductivity and type-II superconducting behavior consistent with the DCM result.

According to Kudo {\it et al.} \cite{Kudo2016}, near the composition where the monoclinic-to-trigonal phase transition occurs in Au$_x$Pd$_{{\rm 1}-x}$Te$_2$, a heat capacity jump at the critical temperature, $\Delta c/\gamma T_{\rm c}$, of about 1.9 was observed due to intermediate-coupling superconductivity. We therefore compared $c_{\rm el}(T)$ with the $\alpha$-model for intermediate-coupling superconductivity using the numerical data tabulated by Johnson \cite{Johnson2013}. The $\alpha$-value can be estimated from the relation: $\Delta c/\gamma T_{\rm c}=\Delta c/\gamma T_{\rm c}|_{\rm BCS}\times(\alpha/\alpha_{\rm BCS})^2$, where $\Delta c/\gamma T_{\rm c}|_{\rm BCS}=1.426$ and $\alpha_{\rm BCS}=1.764$ are the weak-coupling BCS values \cite{Johnson2013}. Using this formula, we calculate the $\alpha$-value for $x=0.35$. At zero field we obtain from Fig.~\ref{fig:3}(d) $\Delta c$ = 0.061 J/Kmol as an idealized step value and $T_{\rm c}$ = 4.6 K as the mid-point of the transition. With the $\gamma$-value given above we calculate $\Delta c/\gamma T_{\rm c} = 1.74$ and consequently $\alpha=1.94$. As shown in Fig.~\ref{fig:3}(d), the measured specific in the superconducting state at zero field agrees well with the numerical data for $\alpha=2~(\sim1.94)$ listed in Ref.\cite{Johnson2013}.

\begin{figure}[htpb]
  \begin{center}
    \includegraphics[clip,width=7.5cm]{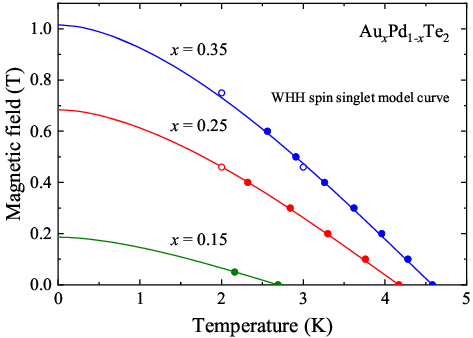}
    \caption{Ambient pressure superconducting phase diagram, $B_{\rm c2}(T)$, of Au$_x$Pd$_{{\rm 1}-x}$Te$_2$ for $x=0.15$, 0.25 and 0.35. Closed and open circles are taken from the HC and DCM measurements, respectively. Solid lines are WHH model curves.}
    \label{fig:4}
  \end{center}
\end{figure}

The normal and superconducting state parameters derived from the HC data for $x=0.35$ are listed in Table~\ref{tab:2}, together with those for $x=0.15$ and 0.25. Regarding $T_{\rm c}$, $\gamma$ and ${\it \Theta}_{\rm D}$, their values and composition variation are consistent with previous research \cite{Kudo2016}. Figure~\ref{fig:4} shows the superconducting phase diagram, obtained from the HC and DCM measurements, for the different $x$-values at ambient pressure. Fitting of $H_{\rm c2}(T)$ with the Werthamer--Helfand--Hohenberg (WHH) model \cite{WHH1996II,WHH1996III} yields the critical field at 0 K: $\mu_0H_{\rm c2}(0)= 0.186$, 0.684 and 1.02 T for $x=0.15, 0.25$ and 0.35, respectively. The coherence length $\xi$ is calculated from the equation $\mu_0H_{\rm c2}(0)=B_{\rm c2}(0)=\Phi_0/(2\pi \xi^2)$, where $\Phi_0$ is the magnetic flux quantum. We estimate $\xi=42.1$, 21.9 and 18.0 nm for $x=0.15, 0.25$ and 0.35, respectively. These $\xi$-values are much smaller than those of the type-I superconductor PdTe$_2$ with $\xi_0=1.8$ $\mu$m for the BCS coherence length and $\xi=439$ nm derived from Ginzburg-Landau (GL) relations \cite{Salis2018}.

\subsection{Pressure and composition dependence of superconducting properties}

\begin{figure}[htpb]
  \begin{center}
    \includegraphics[clip,width=7.5cm]{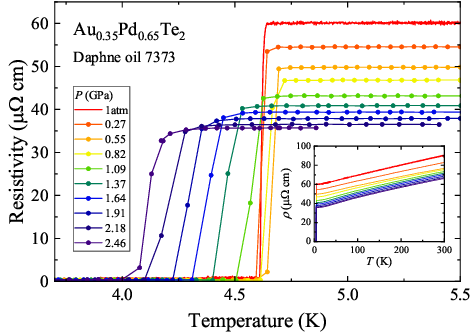}
    \caption{Pressure variation of the superconducting transition of Au$_{0.35}$Pd$_{0.65}$Te$_2$ in the electrical resistivity. The inset shows $\rho(T)$ up to room temperature with increasing pressure. Color variations related to the experimental pressure are the same in the main frame and the inset.}
    \label{fig:5}
  \end{center}
\end{figure}

Figure~\ref{fig:5} shows the temperature variation of the electrical resistivity, $\rho(T)$, at applied pressures up to 2.46 GPa for $x = 0.35$, measured in zero-field; the main panel presents a magnified view around the transition. With increasing pressure up to 0.82 GPa superconductivity is enhanced, and successively depressed with a slight broadening of the transition width. As shown in the inset of Fig.~\ref{fig:5}, $\rho(T)$ exhibits metallic behavior, and its slope remains mostly unchanged, indicating that the electronic state varies monotonically with pressure. 

\begin{figure}[htpb]
  \begin{center}
    \includegraphics[clip,width=6.0cm]{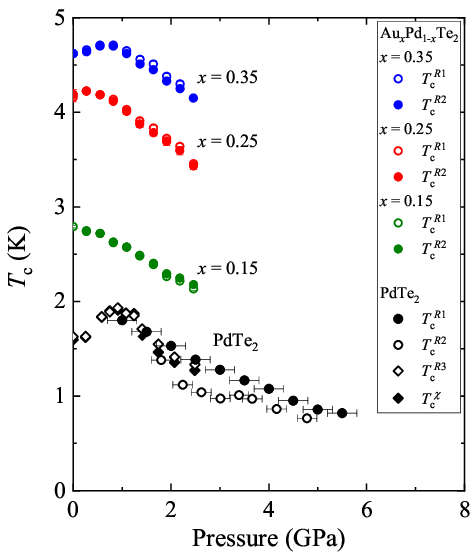}
    \caption{Superconducting transition temperatures, $T_{\rm c}$, as a function of pressure for Au$_x$Pd$_{{\rm 1}-x}$Te$_2$ for different $x$-values as indicated. Open and closed circles indicate values obtained on samples R1 and R2, respectively. For comparison, $T_{\rm c}(P)$ of PdTe$_2$, obtained from our previous resistivity and ac-susceptibility measurements, is also plotted \cite{Leng2020, Furue2021}.}
    \label{fig:6}
  \end{center}
\end{figure}

The value of $T_{\rm c}$ was determined in the same way as shown in the inset of Fig.~\ref{fig:3}(a). In Figure~\ref{fig:6} we trace the pressure variation, $T_{\rm c}(P)$, for each composition of the Au substituted compounds, together with the one of PdTe$_2$ for comparison \cite{Leng2020, Furue2021}. The data points for each composition were obtained by measuring two samples (labeled R1 and R2) and provide good reproducibility, showing homogeneity in the synthesized samples. For $x=0.25$ and 0.35 $T_{\rm c}(P)$ passes through a shallow maximum, with $T_{\rm c}^{\rm max}$= 4.2 K at $P=0.27$ GPa and 4.7 K at $P= 0.55-0.82$ GPa, respectively. Compared to PdTe$_2$, $T_{\rm c}^{\rm max}$ appears at a lower pressure, and the increment between $T_{\rm c}(0)$ and $T_{\rm c}^{\rm max}$ is small. In the doped compounds, the pressure at which $T_{\rm c}$ reaches its maximum shifts to lower values with a smaller $x$-value.

\begin{figure}[htbp]
  \begin{center}
    \includegraphics[clip,width=7.5cm]{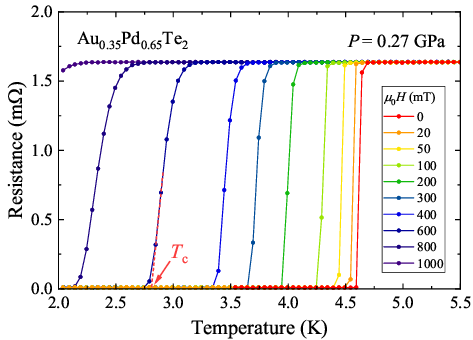}
    \caption{Electrical resistances of Au$_{0.35}$Pd$_{0.65}$Te$_2$ at 0.27 GPa as a function of temperature under the applied magnetic fields from 0 to 1T.}
    \label{fig:7}
  \end{center}
\end{figure}

\begin{figure}[htbp]
  \begin{center}
    \includegraphics[clip,width=7.5cm]{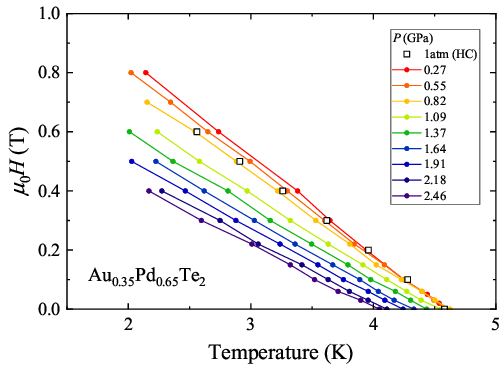}
    \caption{Pressure variation of $H_{\rm c2}$ in the range 0.27 - 2.46 GPa as indicated by different colors for Au$_{0.35}$Pd$_{0.65}$Te$_2$. Open squares are data points at 1 atm obtained from the heat capacity (HC) measurements.}
    \label{fig:8}
  \end{center}
\end{figure}

Next, we investigated at each pressure the depression of superconductivity by a magnetic field. In Figure~\ref{fig:7} we show a typical resistance data set taken at 0.27 GPa for $x=0.35$. The transition is gradually depressed with increasing field and shifts to below 2 K in a field of 1 T. The value of $T_{\rm c}$ was determined at each field by the intersection of the extrapolated $R(T)$ curve at the transition and the zero resistance axis. In Figure~\ref{fig:8} we plot the $T_{\rm c}$-values obtained for $x=0.35$, and construct the $H_{\rm c2}(T)$ curve at each pressure. With increasing pressure, the overall slope $-dH_{\rm c2}(T)/dT$ decreases, indicating that $H_{\rm c2}(0)$ becomes smaller with a decrease in $T_{\rm c}$. 

\begin{figure}[htbp]
  \begin{center}
    \includegraphics[clip,width=8.5cm]{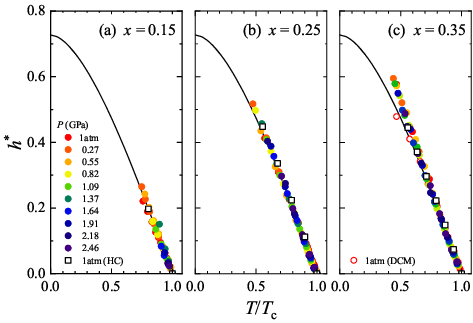}
    \caption{Pressure data collapse of the reduced critical field plots, $h^*=(H_{\rm c2}(T)/T_{\rm c})/(-dH_{\rm c2}/dT)|_{T_{\rm c}}$ versus $T/T_{\rm c}$, for Au$_x$Pd$_{{\rm 1}-x}$Te$_2$ with (a) $x=0.15$, (b) $x=0.25$, and (c) $x=0.35$. Closed circles are taken from resistivity measurements at ambient and high pressure. Open squares are taken from heat capacity (HC) measurements and the open circle in (c) is from DC magnetization (DCM) at 1 atm. Solid lines in each panel represent the WHH model curve in the clean limit.}
    \label{fig:9}
  \end{center}
\end{figure}

In Figure~\ref{fig:9} we show the superconducting phase diagram at different pressures in a reduced form, $h^*=(H_{\rm c2}(T)/T_{\rm c})/(-dH_{\rm c2}/dT)|_{T_{\rm c}}$ versus $T/T_{\rm c}$. The curves at different pressures collapse on a single curve, implying that the superconducting properties are unchanged throughout the pressure range up to $\sim2.5$ GPa. Moreover, the collapsed curves agree well with the WHH model curve. For $x=0.35$ the $h^*$ plot departs somewhat below $T/T_{\rm c}\sim 0.6$, when determined from the resistance data. Since the measurements are made on polycrystalline samples and resistance is not a bulk probe, we do not think this is significant. Notice, $H_{\rm c2}$ determined by specific heat, which is a bulk probe, is in agreement with the model curve.

No signatures of surface superconductivity were observed within the resolution of our measurements. We remark that in PdTe$_2$ crystals, surface superconductivity was detected by ac-susceptibility measurements in small driving fields, and in the resistance data in applied magnetic fields by the critical field being enhanced over that of the bulk \cite{Leng2017}. The resistance data on our polycrystalline doped compounds do not show such an enhancement. Possibly, grain boundary scattering or surface disorder inherent in polycrystalline samples impedes surface superconductivity. Notice, in Pt doped PdTe$_2$ crystals signatures of surface superconductivity were reported, although not as pronounced as in pure PdTe$_2$ \cite{Salis2022}.

\subsection{Electronic structure calculations at ambient pressure} 

\begin{figure}[t]
  \begin{center}
    \includegraphics[clip,width=7.4cm]{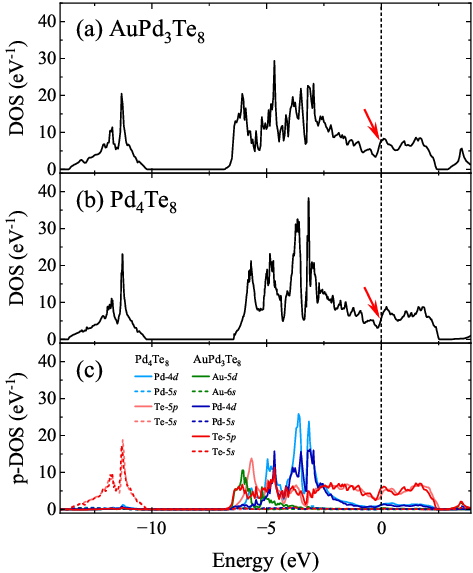}
    \caption{he total DOS per supercell of (a) AuPd$_3$Te$_8$ and (b) Pd$_4$Te$_8$, which are calculated based on the structural model with a 2$\times$2$\times$1 supercell (see text), and (c) their partial DOS. The vertical broken line and red arrows indicate the Fermi energy and $N(E_f)$ in each composition, respectively. Notice that for AuPd$_3$T$_8$ $E_{f}$ shifts to close to the maximum of a pronounced DOS peak (panel (a)).}
    \label{fig:10}
  \end{center}
\end{figure}

We here calculate the DOS of two structural models to investigate the effect of Au substitution on the electronic structure. One is a 2$\times$2$\times$1 supercell of pure PdTe$_2$ with the structural parameters taken from Ref. \cite{Pell1996} (referred to as Pd$_4$Te$_8$ hereafter). The other is a 2$\times$2$\times$1 supercell of the Au substituted PdTe$_2$, Au$_{0.25}$Pd$_{0.75}$Te$_2$, in which one Pd atom was replaced by one Au atom (referred to as AuPd$_3$Te$_8$ hereafter). The latter model uses the present experimental lattice parameters of Au$_{0.25}$Pd$_{0.75}$Te$_2$ determined in this work. The details of the structural models calculations are presented in Sect. 6 of the SM. Figure \ref{fig:10}(a) and (b) display the total DOS of AuPd$_3$Te$_8$ and Pd$_4$Te$_8$, respectively. The overall shape of the DOS for AuPd$_3$Te$_8$ is similar to that of Pd$_4$Te$_8$, although the DOS peaks are broader in the Au substituted compound. The values of $N(E_f)$ in units of states /eV per supercell increases from 5.6 for Pd$_4$Te$_8$ to 7.8 for AuPd$_3$Te$_8$. $E_f$ shifts to a higher energy, which is in line with AuPd$_3$Te$_8$ having one more valence electron than Pd$_4$Te$_8$ (see Sect. 6 in the SM). Furthermore, the dominant contribution to $N(E_f)$ is due to the Te-5$p$ orbital for both compounds, as shown in Fig. \ref{fig:10}(c). 


\section{Discussion}

\subsection{Composition dependence of superconductivity at ambient pressure}

The experimental results show that superconductivity of the Au substituted PdTe$_2$ compounds has a type-II nature. We first evaluate the basic superconducting parameters at ambient pressure. In Table~\ref{tab:3} the values of the relevant parameters obtained from the HC and DCM data are listed, together with $H_{\rm c2}$ and $\xi$ values taken from Table~\ref{tab:2}. The thermodynamic critical field $H_{\rm c}$ was obtained from the HC data using the relation $\Delta c|_{T_{\rm c}} = 4[\mu_0H_{\rm c}(0)]^2/\mu_0 T_{\rm c}$, where $c$ is in units of J/m$^3$. The GL parameters, which are calculated from the relation $\kappa = H_{\rm c2}/(\sqrt{2}H_{\rm c})$, are larger than the type-I/II threshold $\kappa=1/\sqrt{2}$, thereby confirming type-II superconductivity. Alternatively, $H_{\rm c}$ can also be determined from the DCM data using the equation $-\mu_0 \int M dH=\mu_0 H_{\rm c}^2/2$, where $M$ is the DC magnetization from 0 to $H_{\rm c2}$ measured with increasing applied field. Regarding the difference in the $\kappa$-values obtained from the two experiments, those from the HC data are probably more reliable, since the actual $M$--$H$ curve is rounded due to sample geometry, which might lead to an underestimation of the $\kappa$-value. Finally, we list in Table~\ref{tab:3} values of the magnetic penetration depth, calculated from the relation $\lambda = \kappa \xi$.


The superconducting transition temperature $T_{\rm c}$ increases with Au content. Within the McMillan approach, $T_{\rm c}$ is given by multiplying the ${\it \Theta}_{\rm D}$-coefficient with an exponential function composed of an electron-phonon (e-ph) coupling constant, $\lambda_{\rm ep}$, and the Coulomb pseudopotential, $\mu_{\rm c}^*$ \cite{McMillan1968}. With the measured values of $T_{\rm c}$ and ${\it \Theta}_{\rm D}$ taken from Table II, and a typical value for $\mu_{\rm c}^*$ of 0.1 \cite{McMillan1968}, we calculated $\lambda_{\rm ep}$ is 0.53, 0.62 and 0.66 for $x=0.15$, 0.25 and 0.35, respectively. This is in-line with an increase of the measured electronic coefficient in the specific heat $\gamma = \gamma_0(1+\lambda_{\rm ep})$, where $\gamma_0$ is the bare $\gamma$ value (Table II). We remark that the calculated DOS at $E_f$, $N(E_f)$, which is proportional to $\gamma_0$, also increases from 5.6 states/eV for Pd$_4$Te$_8$ to 7.8 states/eV for AuPd$_3$Te$_8$. Using the experimental $\gamma$ values of 4.4 mJ/mol K$^2$ for PdTe$_2$ \cite{Salis2021} and 7.15 mJ/mol K$^2$ for $x=0.25$ (Table II) we calculate $\lambda_{\rm ep}$ is 0.33 and 0.56, respectively. The increase of $\lambda_{\rm ep}$ with Au substitution complies with the trend obtained from the McMillan equation.




\begin{table}[htpb]
\caption{\label{tab:3} The superconducting parameters obtained from the HC and DCM measurements: the upper critical field $\mu_0 H_{\rm c2}$ (mT), the thermodynamic critical field $\mu_0 H_{\rm c}$ (mT), the GL parameter $\kappa$, the coherence length $\xi$ (nm), and the magnetic penetration depth $\lambda$ (nm). }
\begin{ruledtabular}
\begin{tabular}{cccccc}
\mbox{$x$}&\mbox{$\mu_0 H_{\rm c2}$}&\mbox{$\mu_0 H_{\rm c}$}&\mbox{$\kappa$}&\mbox{$\xi$}&\mbox{$\lambda$}\\
\mbox{}&\mbox{(mT)}&\mbox{(mT)}&\mbox{}&\mbox{(nm)}&\mbox{(nm)}\\
\hline\vspace{-1mm}\\
Heat capacity &&&&& \vspace{1mm}\\ 
0.15  &186& 35.2 & 3.7 & 42.1  & 155  \\
0.25  &684&80.2 & 6.0 & 21.9  & 131   \\
0.35 &1016& 95.5 & 7.5  & 18.0 & 135   \\
\hline\vspace{-1mm}\\
dc mag. ($T=2$ K) &&&&& \vspace{1mm}\\
0.25  &460&70.7 & 4.6 & 26.7  & 122   \\
0.35 &744& 148 & 3.5  & 21.0 & 73.5   \\
\end{tabular}
\end{ruledtabular}
\end{table}

\subsection{Pressure variation of $T_{\rm c}$ at each composition}

\begin{figure}[htpb]
  \begin{center}
    \includegraphics[clip,width=8.5cm]{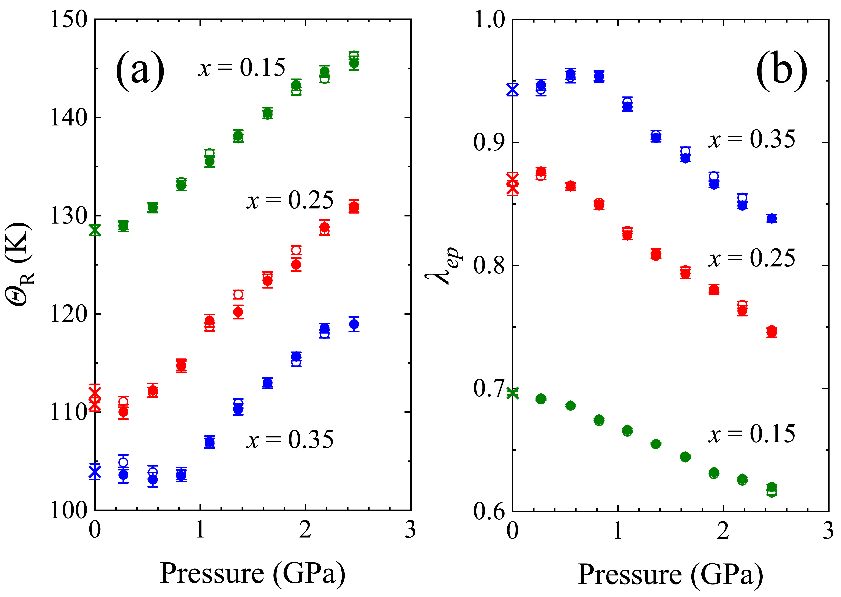 }
    \caption{Pressure variation of (a) the transport Debye temperature ${\it\Theta}_{\rm R}$ derived from the $R-T$ data below 100 K, and (b) the effective electron-phonon coupling constant $\lambda_{\rm ep}$  calculated from $T_{\rm c}$ and ${\it \Theta}_{\rm R}$, for Au$_x$Pd$_{{\rm 1}-x}$Te$_2$ with different $x$-values as indicated. Open and closed circles are obtained for sample 1 and sample 2, respectively, under pressure. The cross marks are obtained at ambient pressure.}
    \label{fig:11}
  \end{center}
\end{figure}

In general, applying pressure tends to depress superconductivity, whereas $T_{\rm c}(P)$ for $x=0.25$ and 0.35 has a maximum. To discuss $T_{\rm c}(P)$ of the Au substituted compounds, we estimated the pressure variation of $\lambda_{\rm ep}$ based on the McMillan approach as reported in the previous section. However, in stead of using ${\it \Theta}_{\rm D}$ we use the transport Debye temperature, ${\it \Theta}_{\rm R}$ \cite{Ziman1960}, as explained below. The transport Debye temperature, ${\it \Theta}_{\rm R}$, was extracted at each pressure by fitting the resistance data below 100 K to the Bloch-Gr\"{u}neisen formula (see Sect. 7 in the SM). Figure~\ref{fig:11} (a) shows the pressure variation of ${\it \Theta}_{\rm R}(P)$ obtained at each composition. We remark that the ${\it \Theta}_{\rm R}$-values may differ from the thermodynamic ${\it \Theta}_{\rm D}$-values, notably when working with polycrystalline samples, since grain-boundary and defect scattering contribute to the electrical resistivity as well. However, the ${\it \Theta}_{\rm R}$-value also reflects the variation of lattice stiffness and phonon energy due to the change in volume \cite{Ziman1960}. Since we are mainly dealing with the relative variation ($d{\it \Theta}_{\rm R}/dP$) our approach is sound, and we use ${\it \Theta}_{\rm R}$ in stead of ${\it \Theta}_{\rm D}$ in the McMillan formula. 

The pressure variation $\lambda_{\rm ep}(P)$, deduced in this way, is plotted in Fig.~\ref{fig:11}(b). Interestingly, $\lambda_{\rm ep}(P)$ and ${\it \Theta}_{\rm R}(P)$ exhibit a striking opposite variation with pressure at each composition. It implies the  pressure variation $T_{\rm c}(P)$ can be attributed to a competition between the phonon energy and the electron–phonon coupling strength. For $x=0.25$ and 0.35, the slight decrease of ${\it \Theta}_{\rm R}$ indicates initially phonon softening, which tends to decrease $T_{\rm c}$. However, the e-ph coupling is simultaneously enhanced and appears to outweighs this effect, resulting in an increase of $T_{\rm c}(P)$. At higher pressures, the monotonous increase of ${\it \Theta}_{\rm R}$ and decrease of $\lambda_{\rm ep}$ account for the depression of $T_{\rm c}$ (shown in Fig.~\ref{fig:6}). This tells us that the variation of the e-ph coupling plays a dominant role in determining  $T_{\rm c}$. Additionally, the anti-symmetric behavior of ${\it \Theta}_{\rm R}(P)$ and $\lambda_{\rm ep}(P)$ indicates that lattice stiffness, which relates to the phonon frequencies, significantly affects the e-ph coupling and its pressure variation.

In order to shed more light on the non-monotonous variation of $T_{\rm c}$, ${\it \Theta}_{\rm R}$ and $\lambda_{\rm ep}$, we performed a finite-strain analysis \cite{Furue2021}. In brief, the relationship $H_{\rm v}(f_{\rm e})$ between a normalized pressure $H_{\rm v}$ and an Eulerian strain $f_{\rm e}$ is calculated from the lattice parameters. A change in the slope of $H_{\rm v}(f_{\rm e})$ indicates a change in the bulk modulus. This in turn might indicate a change in the topology of the Fermi surface, $\it{e.g.}$, a Lifshitz transition \cite{Lifshitz1960, Polian2011}. We present the finite strain analysis in Sect. 8 of the SM. Fig. S13 shows that the slopes of the $H_{\rm v}(f_{\rm e})$ curves for $x=0.25$ and 0.35 change significantly near $f_{\rm e} \sim 0.003$: negative below this value and positive there above. The changing point corresponds to a pressure of $0.4-0.6$ GPa. For $x = 0.15$ $H_{\rm v}$ remains flat below $f_{\rm e} \sim 0.01$ within the error bar. 

We first discuss whether the slope change for $x=0.25$ and 0.35 near $0.4-0.6$ GPa can be due to a Lifshitz transition, related to a van Hove singularity (vHs). Kim {\it et al.} \cite{Kim2018} performed detailed electronic and phononic structure calculations for PdTe$_2$. A saddle point vHs near the M point was found to be responsible for superconductivity. The calculations show that a pressure of more than 10 GPa is required to bring the Fermi energy $E_{f}$ to the M point and induce a Lifshitz transition. This pressure is larger than pressures we applied in our previous study \cite{Furue2021} and we concluded we could not tune the vHs to the Fermi level in PdTe$_2$. On the other hand, Kim {\it et al.} reported that electron doping results in an upward shift of $E_{f}$: doping of 0.18 electrons per unit cell results in a shift of 0.12 eV, which brings $E_{f}$ to the vHs. Au doping results in an increase of the carrier concentration (see Sect. 2 in the SM). According to our electronic structure calculations for $x=0.25$ (Sect. III-D), doping of 0.25 electron per unit cell induces an upward shift of $E_{f}$ ($\sim 0.14$ eV), with $E_{f}$ located near the maximum of a sharp DOS peak. This could be the effect of the vHs brought closer to $E_{f}$. At the same time, pressure tuning of the vHs through $E_{f}$ becomes feasible within our pressure range. This offers a possible explanation for the maximum in $T_{\rm c}(P)$ near 0.3 and 0.7 GPa for $x=0.25$ and 0.35, respectively. On the other hand, the band structure of the Au substituted compounds is not rigid and will differ in detail from the one of PdTe$_2$, thus one should be careful with this Lifshitz shift model. 

A more straight-forward interpretation of the finite-strain analysis relies on the relationship with ${\it \Theta}_{\rm R}(P)$. The slope change of $H_{\rm v}(f_{\rm e})$ for $x=0.25$ and 0.35 implies that the bulk moduli change to a lower value at high pressures, \textit{i.e.} above $P\sim 0.4-0.6$ GPa. In general, the pressure variation of the Debye temperature is larger in a smaller bulk modulus state. This trend is observed in Fig.~\ref{fig:11}(a). Therefore, the finite-strain analysis complies with the more simple explanation of relating $T_{\rm c}$, ${\it \Theta}_{\rm R}(P)$ and $\lambda_{\rm ep}(P)$ by the McMillan model. The same reasoning holds for PdTe$_2$ \cite{Furue2021}. 

\begin{figure}[t]
  \begin{center}
    \includegraphics[clip,width=8cm]{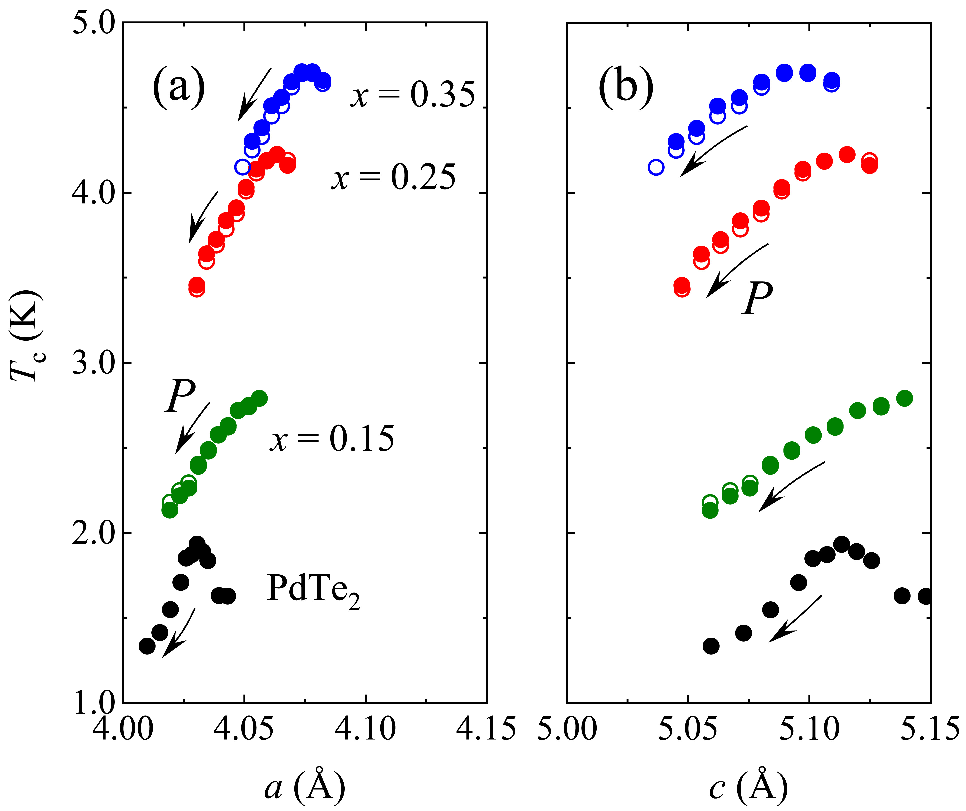}
    \caption{The superconducting transition temperature, $T_{\rm c}$, as a function of lattice constants (a) $a$-axis and (b) $c$-axis. The experimental data for PdTe$_2$ are cited from ref.~\cite{Leng2020, Furue2021}. Arrows indicate the direction of increasing applied pressure at each composition.}
    \label{fig:12}
  \end{center}
\end{figure}

A final interesting observation as regards the pressure dependence of $T_{\rm c}$ can be made when inspecting Fig.~\ref{fig:12}, where we trace $T_{\rm c}$ as a function of the lattice parameters $a$ and $c$. The strong dependence of $T_{\rm c}$ on the $a$ lattice parameter compared to $c$ is in-line with the calculations by Kim \textit{et al.} that predict that the Te-shear $O_{1,2}$ phonon mode (an in-plane vibrational mode) strongly correlates with superconductivity \cite{Kim2018}. 

\subsection{Composition dependence of $T_{\rm c}(P)$} 

\begin{figure}[t]
  \begin{center}
    \includegraphics[clip,width=7cm]{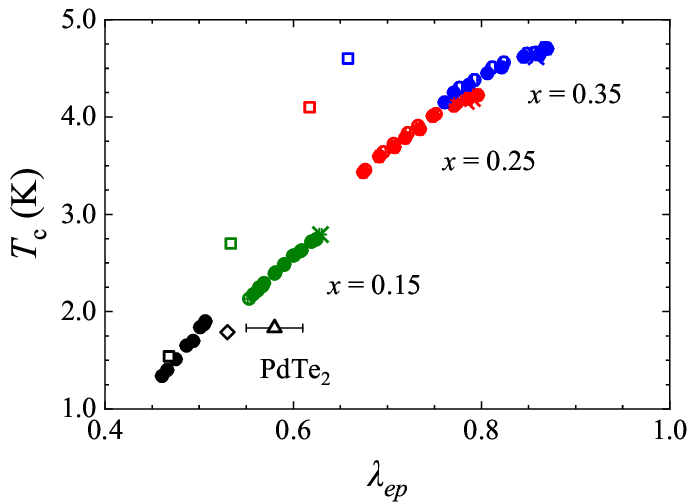}
    \caption{The transition temperature as a function of the electron-phonon coupling constant deduced from the McMillan equation for Au$_x$Pd$_{{\rm 1}-x}$Te$_2$. The open and closed circles are for sample 1 and sample 2, respectively, with the ${\it \Theta}_{\rm R}$-values obtained by fitting the resistance data under pressure. Data points for PdTe$_2$ were recalculated from \cite{Furue2021}. The cross marks are obtained using ambient pressure transport data. The open squares are calculated with ${\it \Theta}_{\rm D}$ obtained from the heat capacity at ambient pressure. The open diamond and triangle show literature values reported in Refs. \cite{Kim2018, Anemone2021}.}
    \label{fig:13}
  \end{center}
\end{figure}

In this section we compare $T_{\rm c}(P)$ obtained for the different Au$_x$Pd$_{{\rm 1}-x}$Te$_2$ compositions. The main results are collected in Fig.~\ref{fig:13} where $T_{\rm c}(P)$ is plotted as a function of $\lambda_{\rm ep}$. The error bars are given by the size of the symbols. Overall, the $T_{\rm c}$ values show a systematic behavior and present a smooth function of $\lambda_{\rm ep}$. This validates our McMillan approach across the different samples. We remark that the $\lambda_{\rm ep}$ values deduced with ${\it \Theta}_{\rm D}$ from the specific heat are considerably lower than those calculated with ${\it \Theta}_{\rm R}$ (see also Sect. IV-B), but importantly show the same trend. We include two data points for PdTe$_2$ obtained from theoretical calculations ($\lambda_{\rm ep}$ = 0.53) \cite{Kim2018} and an experimental value obtained by helium atom scattering ($\lambda_{\rm ep}$ = 0.58$\pm$0.03) \cite{Anemone2021}. Both values coincide.

Next, we consider the composition dependence of $T_{\rm c}$ from a structural viewpoint. Figure~\ref{fig:12}(a) and (b) show the $T_{\rm c}$-values for the substituted compounds and pure PdTe$_2$ as a function of the lattice constants $a$ and $c$, respectively. While the  $a$-axis expands, the $c$-axis shrinks with Au substitution (Table I), which results in a volume increases. This can be seen as a negative chemical pressure. Accordingly, hydrostatic pressure results in an overall decrease of $T_{\rm c}$. In the CdI$_2$-type structure, the sites where Au atoms replace Pd atoms are arranged in the $ab$\,($aa$)-plane perpendicular to the $c$-axis. Given the wide compositional range, in-plane deformation likely plays a more significant role in tuning the electronic states and superconductivity than interlayer deformation.

Finally, we notice a distinct feature when comparing the Au substituted compounds with PdTe$_2$: the intralayer angle Te-Pd-Te in pure PdTe$_2$ shows a weak maximum around 1 GPa, $i.e.$ the pressure where $T_{\rm c}$ has a maximum \cite{Furue2021}. It was suggested that this structural feature was possibly related to the change of the bulk modulus deduced from the finite-strain analysis. In the Au substituted compounds the angle Te-Pd/Au-Te varies monotonically with pressure and does not show an anomalous behavior (see Fig. S6(d) in the SM). 

\section{Summary and conclusions}
 
The series Au$_x$Pd$_{{\rm 1}-x}$Te$_2$ offers an excellent opportunity to study superconductivity in the CdI$_2$-type structure over a large concentration range. We have investigated the effect of pressure on the structural and superconducting properties of polycrystalline samples prepared with $x=0.15$, 0.25 and 0.35, and superconducting transition temperatures $T_{\rm c}=2.7$, 4.1 and 4.6 K, respectively. Synchrotron radiation x-ray diffraction shows that the CdI$_2$-type structure remains stable up to 8 GPa for these compositions. The bulk moduli are obtained by fitting the lattice parameters to the Birch-Murnaghan equation of state and fall in the range $59-63$ GPa.

Superconductivity at ambient pressure was studied by resistance, magnetization and heat capacity. From the heat capacity data we conclude the Au substituted samples are conventional type-II superconductors, albeit that the coupling strength increases from weak for $x=0.15$ to intermediate for $x=0.25$ and 0.35. Employing the McMillan equation with ${\it \Theta}_{\rm D}$ obtained from the specific heat and $T_{\rm c}$ as input parameters, we calculate electron-phonon coupling parameters $\lambda_{\rm ep}$ of 0.53, 0.62 and 0.66, respectively. Electronic structure calculations for Pd$_4$Te$_8$ and AuPd$_3$Te$_4$ reveal that $E_{f}$ shifts to a robust peak in the DOS, which results in an increase of $N(E_{f})$. This in turn is in agreement with the increase of $T_{\rm c}$ with doping.

Superconductivity under pressure up to 2.5 GPa is investigated by electrical transport. We find that $T_{\rm c}(P)$ for $x=0.25$ and 0.35 has a shallow maximum with $T_{\rm c}^{\rm max}$ values of 4.2 and 4.7 K at $P\sim 0.3$ and 0.7 GPa, respectively. A similar result was obtained for PdTe$_2$ \cite{Leng2020}. For $x=0.15$ $T_{\rm c}(P)$ shows a monotonic decrease. Using the McMillan equation, with this time the Debye temperature ${\it \Theta}_{\rm R}$ extracted from the resistance data under pressure as input, we obtain a remarkable smooth variation of $T_{\rm c}$ with the electron-phonon coupling strength $\lambda_{\rm ep}$, spanning the entire pressure (up to 2.5 GPa) and concentration range (from pure PdTe$_2$ to $x=0.35$). From this we conclude superconductivity across the Au$_x$Pd$_{{\rm 1}-x}$Te$_2$ series under pressure is governed by a delicate balance between lattice stiffening and the strength of the electron phonon coupling.  

Interestingly, the Debye temperature ${\it \Theta}_{\rm R}(P)$ has a weak minimum at the pressures where $T_{\rm c}$ shows a maximum for $x=0.25$ and 0.35. By carrying out a finite strain analysis we investigate whether this anomalous behavior is related to a Lifshitz transition. Particular relevant is here the van Hove singularity (vHs) found by electronic structure calculations at 0.12 eV above $E_{f}$ \cite{Kim2018}. Since Au doping adds carriers and raises the Fermi level it is possible the vHs is brought to $E_{f}$. However, electronic structure calculations are required to put this on firm footing. 

Further insight could be obtained from a study of the Fermi surface under pressure via quantum oscillations. This would require the preparation of single crystals with a sufficiently high quantum mobility. This might also shed light on the evolution of the type-II Dirac physics and its possible connection to superconductivity in this attractive Au$_x$Pd$_{{\rm 1}-x}$Te$_2$ series.

\begin{acknowledgments}
This work was partially supported by a Grant-in-Aid for Scientific Research, KAKENHI (No. 20H01851, 24K06937), NIMS Joint Research Hub Program, the research program on Topological Insulators funded by FOM (Dutch Foundation for Fundamental Research on Matter), CCRF in Niigata University, and performed under Proposal No. 2022G103 and 2024G122 of Photon Factory, KEK.
\end{acknowledgments}



\newpage
\renewcommand{\baselinestretch}{1.2}
\setcounter{figure}{0}
\setcounter{table}{0}
\renewcommand*{\thefigure}{S\arabic{figure}}
\renewcommand*{\thetable}{S\arabic{table}}


\begin{widetext}

\noindent \textbf{\large Supplemental Material for ``Effect of pressure on the superconducting properties of Au substituted PdTe$_2$ with the CdI$_2$-type structure''}

\vspace{1.5cm}


\renewcommand*{\thefigure}{S\arabic{figure}}
\renewcommand*{\thetable}{S\arabic{table}}

\noindent \textbf{1. Characterization of the Au$_x$Pd$_{{\rm 1}-x}$Te$_2$ samples at ambient pressure}\\

The Au$_x$Pd$_{{\rm 1}-x}$Te$_2$ polycrystalline samples were investigated in an Electron Probe Micro Analyzer (EPMA) (company SHIMADZU, type EPMA-1720HT) to analyze their compositions. For the nominal compositions $x = 0.15$, 0.25 and 0.35 the average compositions were determined to be Au$_{0.13}$Pd$_{0.88}$Te$_{1.99}$, Au$_{0.29}$Pd$_{0.74}$Te$_{1.97}$ and Au$_{0.38}$Pd$_{0.66}$Te$_{1.96}$, respectively. Additionally, for the $x = 0.35$ sample, we performed a qualitative analysis using a Scanning Electron Microscope (SEM) (company JEOL, type JSM6500F) with Energy Dispersive X-ray Spectroscopy (EDX) to evaluate its homogeneity and composition. In Figure S1 we show EDX maps for the elements Au, Pd, and Te, and the SEM image. The images confirm a homogeneous composition. The average composition obtained from EDX was determined to be Au$_{0.29}$Pd$_{0.59}$Te$_{2.12}$. There is a certain difference in the value of compositions obtained from EPMA and EDX. In general, EPMA has a higher resolution than EDX and is more accurate. It is considered that the difference in composition can be attributed to the analytical method, as well as sample dependence. The superconducting transition temperature, $T_{\rm c}$, shows a linear relationship with the actual amount of Au substituted as shown in Fig. \ref{fig:s2}. However, we remark that in the manuscript we always use the nominal $x$-values.

\begin{figure}[htpb]
  \begin{center}
\includegraphics[clip,width=10cm]{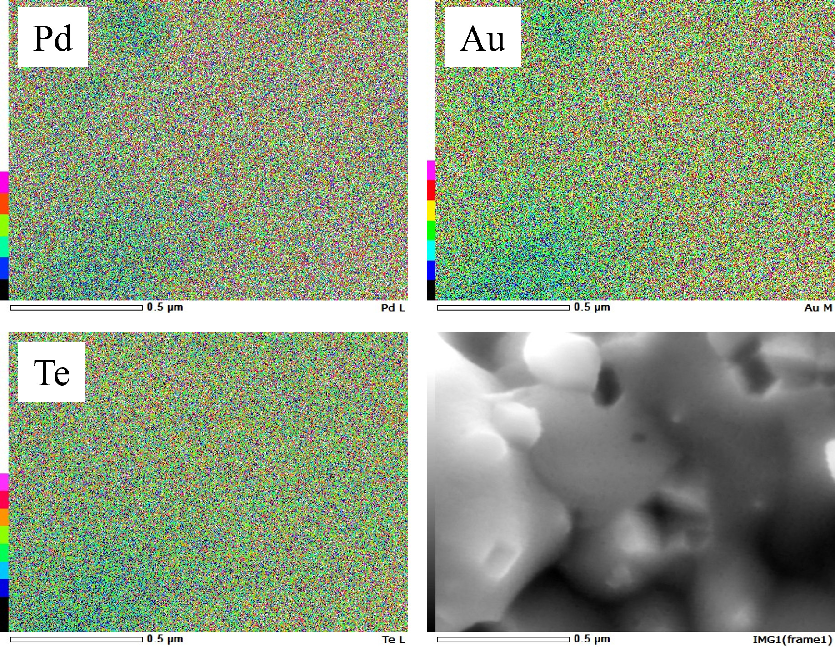}
\caption{EDX images of the elements Pd, Au, and Te, and the SEM image of polycrystalline Au$_{0.35}$Pd$_{0.65}$Te$_2$. The scale bar shown at the bottom left of each image is 0.5 ${\rm \mu}$m.}
\label{fig:s1}
  \end{center}
\end{figure}

\begin{figure}[htpb]
  \begin{center}
\includegraphics[clip,width=8cm]{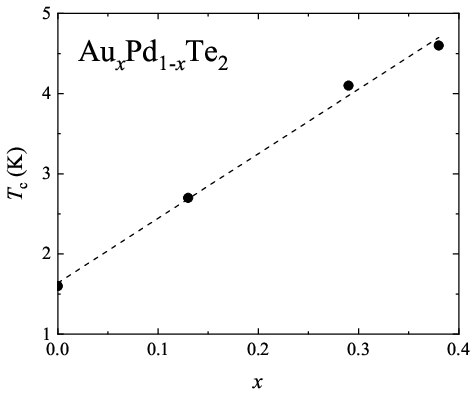}
\caption{The variation of $T_{\rm c}$ as a function of gold concentration $x$.}
\label{fig:s2}
  \end{center}
\end{figure}

\clearpage
\vspace{0.5cm}
\noindent \textbf{2. The Hall effect measurement on the Au$_{0.35}$Pd$_{0.65}$Te$_2$ sample at ambient pressure}\\

In order to determine the carrier concentration at ambient pressure the Hall effect was measured utilizing the Physical Properties Measurement System (Dynacool, Quantum Design). The dimensions  of the polycrystalline Au$_{0.35}$Pd$_{0.65}$Te$_2$ sample were 2.4 (length) $\times$1.1 (width) $\times$0.93 (thickness) mm$^3$. Data were collected with an excitation current $I_{\rm ac} = 10$ mA at $T=50$ K and under magnetic fields up to $B=\pm10$ T. The raw $R_{\rm xy}$-data are shown in Figure~\ref{fig:s3}(a). The parabolic variation is due to a component of the magnetoresistance in $R_{\rm xy}$. After correcting for the magnetoresistance by symmetrizing, the Hall resistance linear in field is obtained as shown in Figure~\ref{fig:s3}(b), and its negative slope indicates that the charge carriers are electrons. The carrier density, $n$, is calculated from the relation $R_{\rm xy}/B=-1/net$, where $e$ is the elementary charge and $t$ the sample thickness, and amounts to $n=1.7\times10^{23}$ cm$^{-3}$. This value of $n$ is one order of magnitude larger than the one of PdTe$_2$: $n=1.1\times10^{22}$ cm$^{-3}$ at $T=50$ K and ambient pressure \cite{Leng2020SM}. We conclude Au doping in PdTe$_2$ strongly increases the carrier density.

\vspace{5mm}
\begin{figure}[h]
  \begin{center}
\includegraphics[clip,width=12cm]{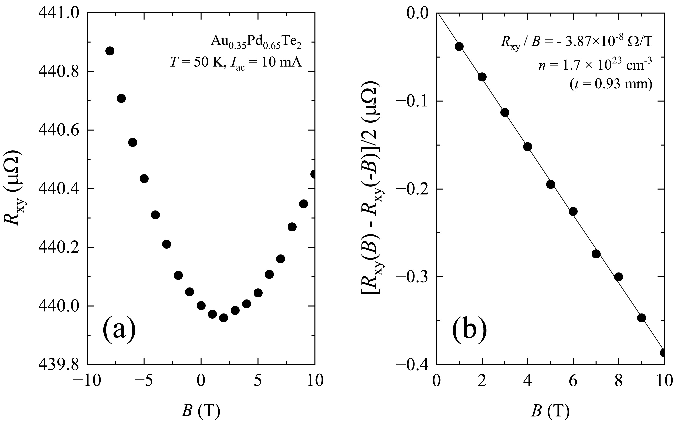}
\caption{(a) Raw Hall resistance as function of magnetic field of polycrystalline Au$_{0.35}$Pd$_{0.65}$Te$_2$. (b) Symmetrized Hall resistance. }
\label{fig:s3}
  \end{center}
\end{figure}

\clearpage
\noindent \textbf{3. Pressure variation of x-ray diffraction patterns of Au$_x$Pd$_{1-x}$Te$_2$ at room temperature and Rietveld refinement}\\

In Figure~\ref{fig:s4} we show the x-ray diffraction patterns of Au$_x$Pd$_{1-x}$Te$_2$ with $x=0.15$, 0.25 and 0.35 measured at room temperature. Most reflections can be indexed with the $P\overline{3}m1$ structure. For each composition, the diffraction pattern profile does not change up to the maximum pressure of $\sim$ 8 GPa, which indicates the $P\overline{3}m$1 structure remains stable in this pressure range. In order to determine the structural parameters, Rietveld analysis was performed using the crystallographic program JANA2020 \cite{Rietveld1967, JANA2006}. In Figure~\ref{fig:s5} we show typical results at low pressure for $x=0.15$ and 0.25; those for $x=0.35$ are shown in the main text. The small residual errors indicates the good quality of the fits. All other diffraction patterns obtained at different pressures were analyzed with a similar accuracy. We remark that the synthesized samples are contaminated with a small amount of gold and/or tellurium dioxide. Additionally, for $x=0.35$, a small peak is detected, that we could not assign, as shown in Fig.~\ref{fig:s4}(c). The diffraction patterns (except the unassigned small peak) for $x=0.35$ were analyzed under condition of the coexistence of the Au$_{0.35}$Pd$_{0.65}$Te$_2$, Au and TeO$_2$ phases. The reliability of the determination of the $z$-position of Te ($z_{\rm Te}$) for $x=0.35$ is somewhat lower than for the other two $x$-values.    

Figure~\ref{fig:s6} shows the deduced structural parameters as a function of pressure: (a) intra Te-Te distance, (b) inter Te-Te distance, (c) intra Pd/Au-Te distance, and (d) intra Te-Pd/Au-Te angle. We remark that there is a discrepancy between the ambient pressure point and the first high-pressure point. We attribute this to the different measuring conditions. The pressure data were measured with the diamond anvil cell (DAC), and the ambient pressure data without a pressure cell. In contrast, the data points for PdTe$_2$ were obtained using a DAC throughout the entire pressure range, and, therefore, the mismatch between the ambient pressure and high pressure data is absent. Table \ref{tab:s1} lists the structural parameters obtained at each pressure and for each composition. The $z_{\rm Te}$-value under high pressure monotonically increases on compression. This variation is different from that in PdTe$_2$: $z_{\rm Te}$ for PdTe$_2$ remains constant within the error up to $\sim$1 GPa and then starts to increase with pressure. 


\begin{figure}[htbp]
  \begin{center}
\includegraphics[clip,width=15cm]{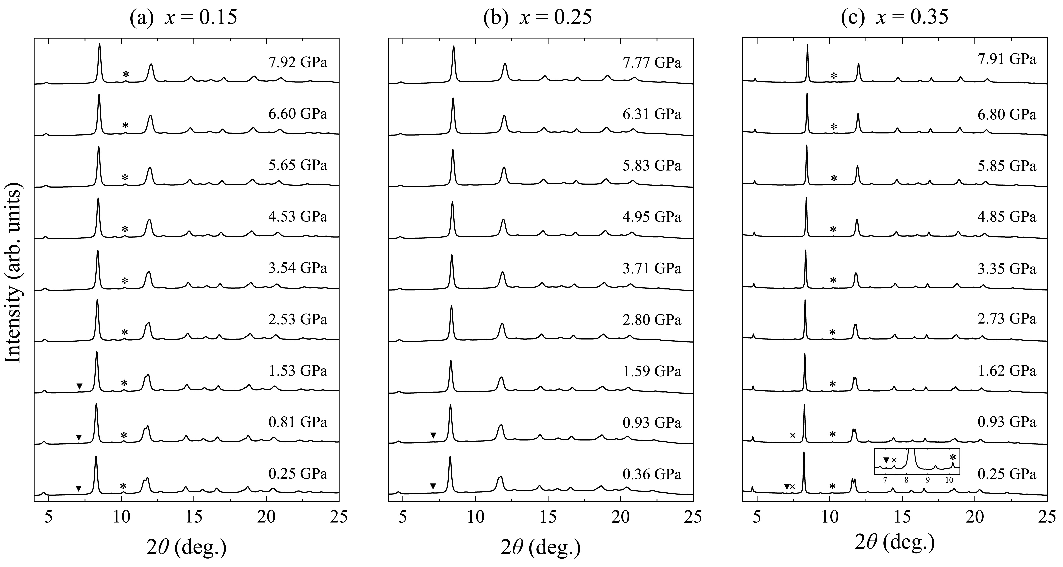}
\caption{Pressure variation of the x-ray diffraction patterns of Au$_x$Pd$_{1-x}$Te$_2$ measured at room temperature: (a) $x=0.15$, (b) $x=0.25$ and (c) $x=0.35$. The asterisk ($\ast$) in (a) and (c) marks the low intensity [111] reflection of gold. The closed triangle ($\blacktriangledown$) indicates the low intensity [110] reflection of TeO$_2$. The cross symbol marks an unassigned peak. The inset at 0.25 GPa in (c) shows a magnified view of the 2$\theta$ range from 6.5 to 10.5 deg.}
\label{fig:s4}
  \end{center}
\end{figure}


\begin{figure}[htbp]
  \begin{center}
\includegraphics[clip,width=15cm]{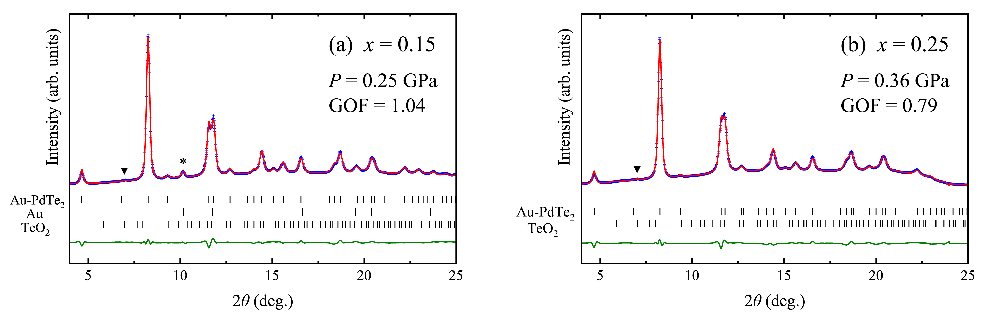}
\caption{Rietveld refinement results of Au$_x$Pd$_{1-x}$Te$_2$: (a) $x=0.15$ at $P=0.25$ GPa, (b) $x=0.25$ at $P=0.36$ GPa. Blue symbols (+), and red and green lines, represent the experimental diffraction intensity, the calculated intensity, and the residual error, respectively. The three sets of vertical bars below the diffraction patterns indicate the diffraction angles of the $P\overline{3}m$1, $Fm\overline{3}m$, and $P4_12_12$ structure for Au$_x$Pd$_{1-x}$Te$_2$, gold ($\ast$ in the pattern), and TeO$_2$ ($\blacktriangledown$ in the pattern), respectively. } 
\label{fig:s5}
  \end{center}
\end{figure}

\begin{figure}[htbp]
  \begin{center}
\includegraphics[clip,width=12cm]{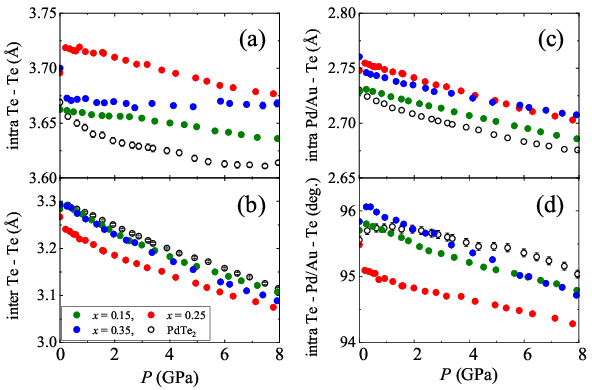}
\caption{Pressure variation of structural parameters: (a) intra Te-Te distance, (b) inter Te-Te distance, (c) intra Pd/Au-Te distance, and (d) intra Te-Pd/Au-Te angle, respectively. Green, red, and blue circles indicate data points for $x=0.15$, 0.25, and 0.35, respectively. The parameters for PdTe$_2$ (open black circles) are plotted for comparison. }
\label{fig:s6}
  \end{center}
\end{figure}

\clearpage
\begin{longtable}{ccccc}

\caption{\label{tab:s1} Structural parameters obtained under high pressure for Au$_x$Pd$_{1-x}$Te$_2$ with $x=0.15$, 0.25 and 0.35: lattice constants $a$(\AA) and $c$(\AA), volume $V$(\AA$^3$) of the unit cell, and the atomic $z$-position of  Te ($z_{\rm Te}$).  In the $P\overline{3}m$1 structure of Au$_x$Pd$_{1-x}$Te$_2$  ($Z$=2), the atomic positions are (0, 0, 0) for Pd/Au and (1/3, 2/3, $z_{\rm Te}$) for Te. As regards the atomic position, only the value of $z_{\rm Te}$ can be optimized by the Rietveld analysis. The asterisk at 1 atm indicates these data were measured without the diamond anvil cell.}\vspace{3mm}\\

\hline \hline Pressure \hspace{5mm} & $a$ & $c$ &  $V$ & $z_{\rm Te}$ \\
(GPa)\hspace{5mm} & (\AA) & (\AA) & (\AA$^3$) & \\

\hline\vspace{1.5mm}
\endfirsthead
\underline{$x=0.15$}\vspace{1mm}\\
1atm$^*$  & 4.0564(3) & 5.1418(8) & 73.263(8)  & 0.2718(5)   \\
0.25 &4.0525(2) & 5.1315(4) & 72.983(5)  & 0.2744(5)  \\
0.49 & 4.0485(2) & 5.1218(4)  & 72.701(5) & 0.2750(5)  \\
0.81  & 4.0440(2) & 5.1118(4)  & 72.397(5) & 0.2757(5)  \\
1.15  & 4.0378(2)  & 5.0980(4) & 71.983(5)  & 0.2764(5)  \\
1.53  & 4.0326(2) & 5.0865(4)  & 71.634(5) & 0.2772(5) \\
1.79  & 4.0286(2)  & 5.0780(4)   & 71.372(4) & 0.2778(5)   \\
2.19 & 4.0221(2)  & 5.0647(4)   & 70.956(4) & 0.2788(5) \\
2.53 & 4.0170(1)  & 5.0542(4)  & 70.630(4) & 0.2796(5)  \\
2.98 & 4.0126(1)  & 5.0444(3)   & 70.340(4) & 0.2801(5)  \\
3.45 & 4.0060(1)  & 5.0315(3)  & 69.926(4) & 0.2808(5) \\
4.00 & 3.9986(1)  & 5.0178(3)  & 69.481(4) & 0.2817(5)  \\
4.53 & 3.9920(1)  & 5.0055(3)  & 69.081(4) & 0.2825(5)  \\
4.98 & 3.9869(1)  & 4.9962(3)  & 68.778(4) & 0.2831(5)  \\
5.65 & 3.9792(1)  & 4.9825(4)  & 68.325(4) & 0.2837(5)  \\
6.14 & 3.9737(1)  & 4.9728(4)  & 68.001(4) & 0.2844(5)  \\
6.60 & 3.9680(1)  & 4.9630(4)  & 67.673(4) & 0.2849(5)  \\
7.19 & 3.9617(2)  & 4.9522(4)  & 67.311(4) & 0.2855(5)  \\
7.92 & 3.9538(2)  & 4.9391(4)  & 66.865(4) & 0.2865(5)  \\

\hline\\
\underline{$x=0.25$}\vspace{1mm}\\
1atm$^*$  & 4.0680(5) & 5.1271(10) & 73.472(10)  & 0.2784(5)   \\
0.07 & 4.0670(2) & 5.1220(4) & 73.371(5)  & 0.2804(5)  \\
0.21 & 4.0652(2) & 5.1190(4)  & 73.262(5) & 0.2817(5)  \\
0.36  & 4.0628(2) & 5.1142(4)  & 73.108(5) & 0.2820(5)  \\
0.50  & 4.0594(2)  & 5.1075(4) & 72.889(5)  & 0.2823(5)  \\
0.58  & 4.0586(2) & 5.1060(4)  & 72.840(5) & 0.2824(5) \\
0.72  & 4.0559(2)  & 5.1000(4)   & 72.655(5) & 0.2833(5)   \\
0.93 & 4.0525(2)  & 5.0937(4)   & 72.444(5) & 0.2833(5) \\
1.22 & 4.0478(2)  & 5.0843(4)  & 72.143(5) & 0.2839(5)  \\
1.55 & 4.0438(2)  & 5.0714(4)   & 71.817(5) & 0.2848(5)  \\
1.59 & 4.0428(2)  & 5.0697(4)  & 71.759(5) & 0.2849(5) \\
1.99 & 4.0368(2)  & 5.0586(4)  & 71.389(5) & 0.2853(5)  \\
2.39 & 4.0301(2)  & 5.0467(4)  & 70.985(5) & 0.2860(5)  \\
2.80 & 4.0250(2)  & 5.0365(4)  & 70.664(5) & 0.2863(5)  \\
3.18 & 4.0207(2)  & 5.0278(4)  & 70.389(4) & 0.2870(5)  \\
3.71 & 4.0148(2)  & 5.0173(4)  & 70.039(4) & 0.2872(5)  \\
4.21 & 4.0065(2)  & 5.0031(4)  & 69.551(4) & 0.2880(5)  \\
4.95 & 3.9981(2)  & 4.9885(4)  & 69.058(4) & 0.2887(5)  \\
5.35 & 3.9917(2)  & 4.9781(4)  & 68.693(4) & 0.2892(5)  \\
5.83 & 3.9848(2)  & 4.9669(4)  & 68.302(4) & 0.2897(5)  \\
6.31 & 3.9789(2)  & 4.9576(4)  & 67.972(4) & 0.2902(5)  \\
7.13 & 3.9706(2)  & 4.9447(4)  & 67.512(4) & 0.2909(5)  \\
7.77 & 3.9628(2)  & 4.9329(4)  & 67.086(4) & 0.2918(5)  \\

\hline\\
\underline{$x=0.35$}\vspace{1mm}\\
1atm$^*$  & 4.0871(4) & 5.1233(9) & 74.100(9)  & 0.2769(5)   \\
0.25 & 4.0834(2) & 5.1110(4) & 73.803(4)  & 0.2755(5)  \\
0.41 & 4.0813(2) & 5.1062(4)  & 73.658(4) & 0.2756(5)  \\
0.63  & 4.0764(2) & 5.0952(4)  & 73.325(4) & 0.2766(5)  \\
0.93  & 4.0715(2)  & 5.0841(4) & 72.987(4)  & 0.2775(5)  \\
1.34  & 4.0653(2) & 5.0703(4)  & 72.569(4) & 0.2781(5) \\
1.62  & 4.0601(2)  & 5.0593(4)   & 72.228(4) & 0.2789(5)   \\
1.96 & 4.0561(2)  & 5.0507(4)   & 71.962(4) & 0.2797(5) \\
2.41 & 4.0496(2)  & 5.0371(4)  & 71.537(4) & 0.2805(5)  \\
2.73 & 4.0452(2)  & 5.0279(4)   & 71.252(4) & 0.2807(5)  \\
3.35 & 4.0370(2)  & 5.0120(4)  & 70.740(4) & 0.2825(5) \\
4.13 & 4.0267(2)  & 4.9921(4)  & 70.098(4) & 0.2839(5)  \\
4.86 & 4.0179(2)  & 4.9757(4)  & 69.563(5) & 0.2851(5)  \\
5.85 & 4.0065(2)  & 4.9559(4)  & 68.894(5) & 0.2874(5)  \\
6.18 & 4.0027(2)  & 4.9495(4)  & 68.674(5) & 0.2877(5)  \\
6.80 & 3.9954(2)  & 4.9375(4)  & 68.258(5) & 0.2887(5)  \\
7.39 & 3.9896(2)  & 4.9284(4)  & 67.936(5) & 0.2894(5)  \\
7.91 & 3.9836(3)  & 4.9197(6)  & 67.612(5) & 0.2904(5)  \\
\hline \hline

\end{longtable}

\clearpage
\noindent \textbf{4. Evaluation of superconducting properties by heat capacity measurement }\\

In Figure~\ref{fig:s7}(a) and (b) we show the temperature dependence of the electronic specific heat, $c_{\rm el} (T )$, under magnetic ﬁeld measured for $x=0.15$ and 0.25, respectively. Similar data for $x=0.35$ are shown in the main text. The phonon contribution was subtracted from the measured specific heat to obtain $c_{\rm el} (T )$ as explained in the main text. By comparing $c_{\rm el} (T )$ to the $\alpha$-model \cite{Johnson2013}, $\alpha$ values of 1.76 and 1.92 $(\sim 2)$ were obtained for $x=0.15$ and 0.25, respectively. This indicates that weak-coupling superconductivity of $x=0.15$ (and PdTe$_2$) evolves into intermediate-coupling for $x=0.25$. The variation of the specific heat as a function of magnetic field is in agreement with these substituted compounds being type-II superconductors. 
 
\begin{figure}[htbp]
  \begin{center}
\includegraphics[clip,width=13cm]{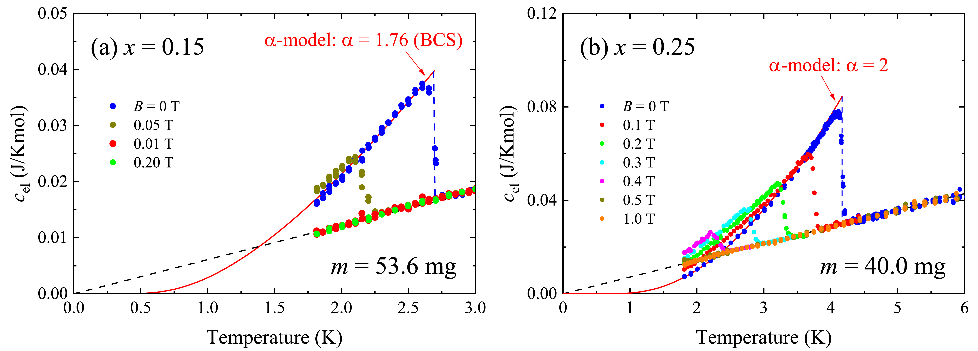}
\caption{Temperature variation of the electronic specific heat of Au$_x$Pd$_{1-x}$Te$_2$ measured in zero field (blue symbols) and increasing magnetic ﬁeld as indicated by the different colors: (a) $x=0.15$, (b) $x=0.25$. The black lines show the normal state linear in $T$ contribution. The solid red lines show the superconducting state specific heat based on the $\alpha$- model. }
\label{fig:s7}
  \end{center}
\end{figure}

\vspace{25mm}
\noindent \textbf{5. Electrical resistivity measurements under high pressure, low temperature, and magnetic field, and superconducting phase diagram}\\

In this section we present the resistance data obtained for $x=0.15$ and 0.25. Resistance data for $x=0.35$ are presented in the main text. Sample sizes for each composition were as follows: $\sim$2.9 (length) $\times$~0.8 (width) $\times$~0.5 (height) mm$^3$ for $x=0.35$, $\sim$1.5${\times}$1.0${\times}$0.2 mm$^3$ for $x=0.25$, and $\sim$2.2${\times}$1.0${\times}$0.2 mm$^3$ for $x=0.15$. Figure~\ref{fig:s8} shows the temperature variation of the electrical resistivity up to a pressure of 2.46 GPa. Resistance measurements under field at each pressure show the depression of superconductivity. Typical data sets measured at 0.27 GPa for $x=0.15$ and 0.25 are shown in Figure~\ref{fig:s9}. The resulting superconducting $H-T$ phase diagrams are shown in Figure~\ref{fig:s10}, where the values of $T_{\rm c}$ were determined as described in Figure 7 in the main text.

\begin{figure}[htbp]
  \begin{center}
\includegraphics[clip,width=13cm]{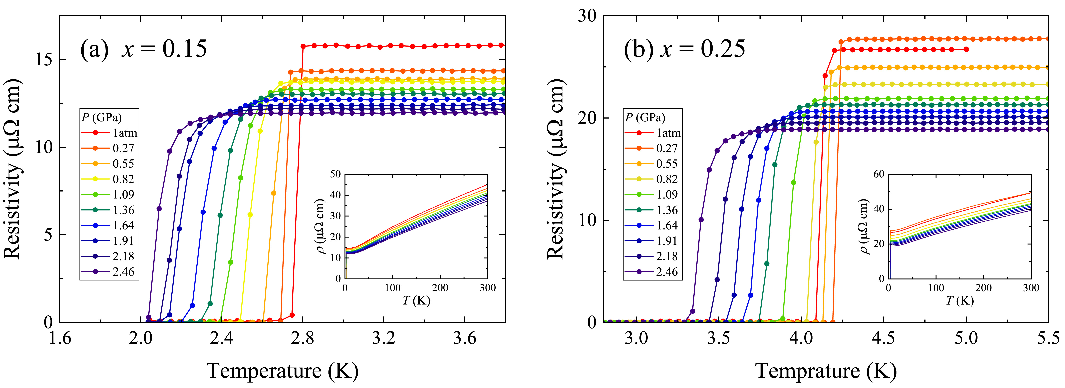}
\caption{Pressure variation of the superconducting transition with increasing pressure: (a) $x=0.15$, (b) $x=0.25$. The inset shows the temperature dependence of the electrical resistivity between 2 and 300 K with increasing pressure. Color variations related to the experimental pressure are the same in both the main and inset ﬁgures.}
\label{fig:s8}
  \end{center}

\vspace{5mm}
  \begin{center}
\includegraphics[clip,width=13cm]{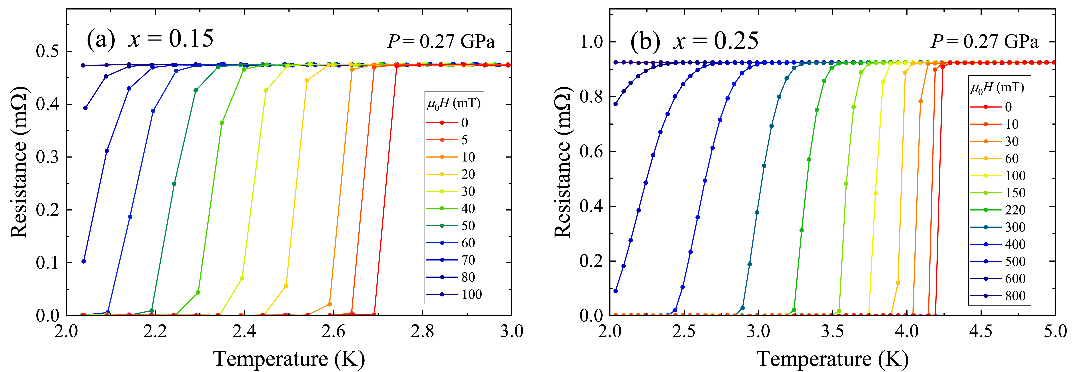}
\caption{Electrical resistance as a function of temperature at 0.27 GPa for (a) $x=0.15$ and (b) $x=0.25$, under applied magnetic ﬁelds up to 100 and 800 mT, respectively.}
\label{fig:s9}
  \end{center}

\vspace{5mm}

  \begin{center}
\includegraphics[clip,width=13cm]{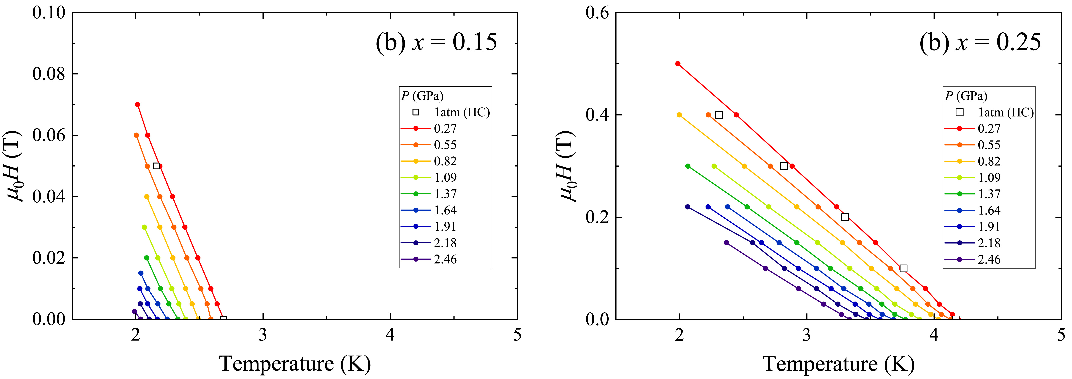}
\caption{Temperature dependence of the critical field $H_{\rm c2}$ at pressures from 0.27 to 2.46 GPa for (a) $x=0.15$ and (b) $x=0.25$. Open squares are data points at 1 atm obtained by heat capacity measurements.}
\label{fig:s10}
  \end{center}
\end{figure}

\clearpage
\noindent \textbf{6. Structural parameters used for calculation of the density of states at ambient pressure.} \\

We show the structural model for the theoretical calculation to obtain the density of states (DOS) at ambient pressure. Table \ref{tab:s2} shows the structural parameters used for the calculation. We calculated the DOS of both the pristine PdTe$_2$ and the substituted material, Au$_{0.25}$Pd$_{0.75}$Te$_2$, with a 2$\times$2$\times$1 supercell. For PdTe$_2$ (Pd$_4$Te$_8$ for the supercell), lattice constants and atomic positions were referred to the values in Ref.\cite{Pell1996}. For the substituted material, Au$_{0.25}$Pd$_{0.75}$Te$_2$ (AuPd$_3$Te$_8$ for the supercell), one Pd atom in the former supercell is replaced by one Au atom, and the lattice constants obtained in this study were used. 

Figure~\ref{fig:s11} illustrates the relation between the total DOS and the integrated DOS (IDOS). The total number of integrated valence electrons for Pd$_4$Te$_8$ and AuPd$_3$Te$_8$ is 88 (10(Pd) + 6(Te)*2, all multiplied by 4) and 89 ([11(Au)+10(Pd)]*3 + 6(Te)*2, all multiplied by 4), respectively. The total DOS for both compounds are almost consistent within the IDOS range of 80-100 around the Fermi energy, indicating that the difference at the $N(E_f)$ is primarily due to a shift of the Fermi energy. 

\vspace{10mm}

\setlength{\tabcolsep}{10pt}
\begin{longtable}{lccc}
\caption{\label{tab:s2} Structural parameters used to calculate the density of states for the two structural models, which are composed of a 2$\times$2$\times$1 supercell. }\vspace{5mm}\\

\hline \hline Lattice constants && $a$ & $c$  \\
for a 2$\times$2$\times$1 supercell&& (\AA) & (\AA)\\
\hline\vspace{0.5mm}\\
Pd$_4$Te$_8$ \cite{Pell1996} && 8.048 & 5.113\\
AuPd$_3$Te$_8$ && 8.136 & 5.127\\
\hline\vspace{5mm}\\

\hline \hline Atomic positions \cite{Pell1996} & $x$ &$y$&$z$  \\
\hline\vspace{0.5mm}\\
Pd1 / Au1 &  0 & 0 & 0\\ 
Pd2 &  0 & 0.5 & 0\\
Pd3 &  0.5 & 0 & 0\\
Pd4 &  0.5 & 0.5 & 0\\
Te1 &   0.16666 &  0.33333  &  0.26628\\
Te2 &   0.16666 &  0.83333  &  0.26628\\
Te3 &   0.66666 &  0.33333  &  0.26628\\
Te4 &   0.66666 &  0.83333  &  0.26628\\
Te5 &   0.33333 &  0.16666  &  0.73372\\
Te6 &   0.33333 &  0.66666  &  0.73372\\
Te7 &   0.83333 &  0.16666  &  0.73372\\
Te8 &   0.83333 &  0.66666  &  0.73372\\
\hline\vspace{7mm}\\
\end{longtable}

\begin{figure}[H]
  \begin{center}
    \includegraphics[clip,width=8.5cm]{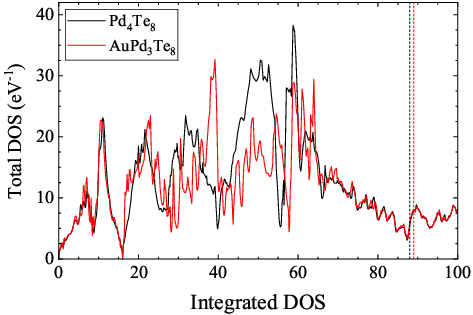}
    \caption{The relation between the total DOS and the integrated DOS of AuPd$_3$Te$_8$ (red) and Pd$_4$Te$_8$ (black). Broken lines indicate the total number of integrated valence electrons in a 2$\times$2$\times$1 supercell,  which corresponds to the Fermi energy.}
    \label{fig:s11}
  \end{center}
\end{figure}

\clearpage
\noindent \textbf{7. Estimation of the transport Debye temperature ${\it \Theta_{\rm R}}$ using the Bloch-Gr\"{u}neisen formula} \\

The pressure variation of the transport Debye temperature, ${\it \Theta_{\rm R}}$, was estimated with help of the Bloch-Gr\"{u}neisen (BG) formula. The $\rho(T)$ data can be fitted to the BG formula based on electron-phonon scattering with a temperature exponent $n$ = 5:
\begin{gather}
\rho(T) =\rho_0 + 4.225 \rho_{\it \Theta} \left (\frac{T}{\it \Theta_{\rm R}}\right)^5 \int_0^{{\it \Theta_{\rm R}}/T} \frac {x^5}{(e^x-1)(1-e^{-x})}dx. \nonumber
\end{gather}
Here $\rho_0$ is the residual resistivity, ${\it \Theta}_{\rm R}$ is the transport Debye temperature, and $\rho_{\it \Theta}$ is the resistivity at ${\it \Theta}_{\rm R}$. Figure~\ref{fig:s12} shows the ﬁtting results for the data obtained at 0.27 GPa for each composition. The temperature range of the fits was limited to below 100 K, which ensures a close to constant pressure. In Figure~\ref{fig:s12} the ${\it \Theta_{\rm R}}$-values obtained at 0.27 GPa are 129, 111 and 105 K for $x=0.15$, 0.25 and 0.35, respectively. We remark the deduced values did not depend on the temperature range below 100 K used for the fitting. Also the results obtained for the two different samples of each composition were in good agreement.  


\vspace{0.5cm}
\begin{figure}[h]
  \begin{center}
\includegraphics[clip,width=8cm]{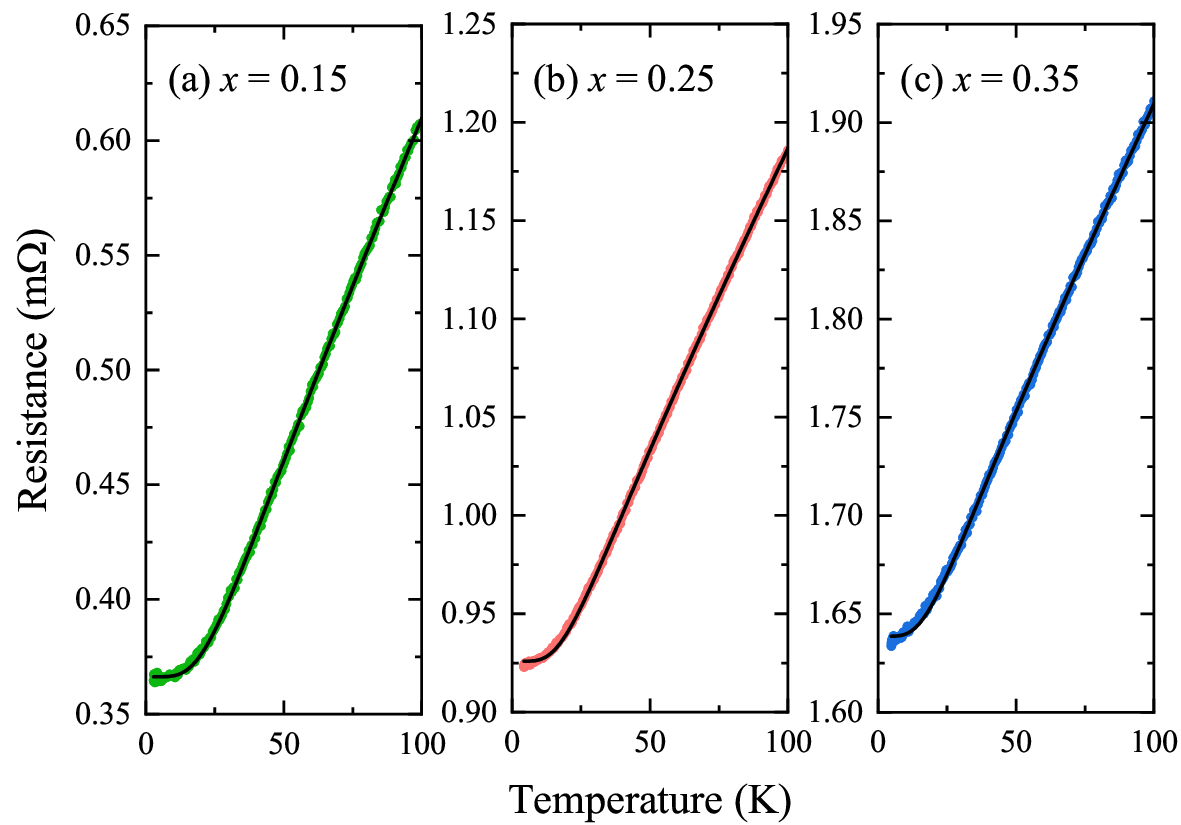}
\caption{Bloch-Gr\"{u}neisen function fits of the resistance of Au$_x$Pd$_{1-x}$Te$_2$ at a pressure of 0.27 GPa for (a) $x=0.15$, (b) $x=0.25$, and (c) $x=0.35$. The colored symbols and black lines indicate the experimental data and fitting results, respectively.}
\label{fig:s12}
  \end{center}
\end{figure}

\clearpage
\noindent \textbf{8. Strain analysis to evaluate fine variations of elasticity}\\

In the same way as for PdTe$_2$ \cite{Furue2021}, we performed a strain analysis using the lattice parameters for each composition. A normalized pressure $H_{\rm v}$ versus an Eulerian strain $f_{\rm e}$ was calculated by applying the experimental lattice constants to the following formula,
\begin{gather}
H_{\rm v} =\frac{P}{3f_{\rm e} (1+2f_{\rm e})^\frac{5}{2}}, ~~~f_{\rm e}=\frac{1}{2}\left [ \left( \frac{V_0}{V}\right)^\frac{2}{3} -1 \right ] \nonumber
\end{gather}
where $V_0$ and $V$ are the volume at ambient and high pressures, respectively,  $P$ is in units of GPa, and
\begin{gather}
H_{\rm v}=B_0 +\frac{3}{2}B_0 (B_0^{'}-4)f_{\rm e}, \nonumber
\end{gather}
where $B_0$ is the bulk modulus, and $B^{'}_0$ its pressure derivative. As can be seen from these formula's, $H_{\rm v}$ and $f_{\rm e}$ have a linear relationship via $B_0$ and $B^{'}_0$. A change in the slope of $H_{\rm v}(f_{\rm e})$ means a subtle change in the bulk modulus. The intercept of the fitted straight line in a specific strain interval and the vertical axis ($f_{\rm e}=0$) provides an estimate of the bulk modulus in each strain region. A change in the slope of $H_{\rm v}(f_{\rm e})$ is often discussed in relation to a change in the topology of the Fermi surface, $\it{e.g.}$ due to a Lifshitz transition \cite{Lifshitz1960, Polian2011}. 

The analytical results for Au$_x$Pd$_{{\rm 1}-x}$Te$_2$ are shown in Figure~\ref{fig:s13}. The error of $H_{\rm v}$ was calculated by propagation of uncertainty. The slopes of $H_{\rm v}(f_{\rm e})$ for $x=0.25$ and 0.35 deviate from the lines extrapolated from the higher $f_e$ range and bend upward below $f_{\rm e}\sim 0.003$, which corresponds to $P\sim 0.4-0.6$ GPa. In contrast, the slope for $x=0.15$ remains approximately constant; the $H_{\rm v}$ values below $f_{\rm e}\sim$ 0.01 fluctuate around the extrapolated line.


\vspace{0.5cm}
\begin{figure}[h]
  \begin{center}
    \includegraphics[clip,width=8.5cm]{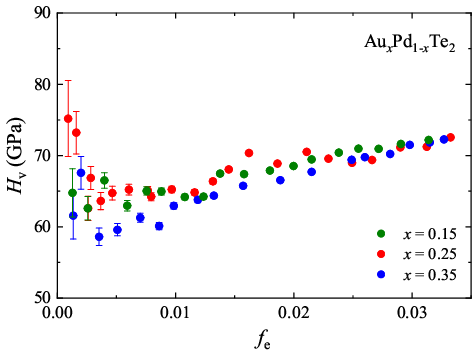}
    \caption{Plot of the normalized pressure $H_{\rm v}$ versus the Eulerian strain $f_{\rm e}$ at ambient temperature for $x=0.15$ (green), 0.25 (red), and 0.35 (blue).}
    \label{fig:s13}
  \end{center}
\end{figure}





\end{widetext}

\end{document}